\documentclass[lettersize,journal]{IEEEtran}

\usepackage{algorithm}
\usepackage{algpseudocode}
\usepackage{amsfonts}
\usepackage{amsmath}
\usepackage{amssymb}
\usepackage{amstext}
\usepackage{array}
\usepackage{balance}
\usepackage{cases}
\usepackage{cite}
\usepackage{cuted}
\usepackage{color}
\usepackage{draftwatermark}
\usepackage{enumerate}
\usepackage{epsfig}
\usepackage{epstopdf}
\usepackage[mathscr]{euscript}
\usepackage{fancyhdr}
\usepackage{footmisc}
\usepackage{gensymb}
\usepackage{graphicx}
\usepackage{latexsym}
\usepackage{longtable}
\usepackage{multirow}
\usepackage{psfrag}
\usepackage{soul}
\usepackage{subcaption}
\usepackage{subeqnarray}
\usepackage{threeparttable}
\usepackage{tikz}
\usetikzlibrary{shapes.geometric, arrows}
\usepackage{wrapfig}

\algblockdefx[Foreach]{BeginForeach}{EndForeach}%
[2]{\textbf{for each} #1 \textbf{in} #2 \textbf{do}}%
{\textbf{end for}}
\algblockdefx[State]{BeginState}{EndState}%
[1]{\textbf{state} #1 \{}%
{\}}
\algblockdefx[For]{BeginFor}{EndFor}%
[3]{\textbf{for} #1 $\gets$ #2 \textbf{to} #3 \textbf{do}}%
{\textbf{end for}}

\newtheorem{remark}{Remark}

\newcommand{\blue}{}

\def\BibTeX{{\rm B\kern-.05em{\sc i\kern-.025em b}\kern-.08em
    T\kern-.1667em\lower.7ex\hbox{E}\kern-.125emX}}

\begin{document}
\title{Efficient Rotating Synthetic Aperture Radar Imaging via Robust Sparse Array Synthesis}
\author{Wei~Zhao,~\IEEEmembership{Student~member,~IEEE},
		Cai~Wen,~\IEEEmembership{Member,~IEEE},
        Quan~Yuan,~\IEEEmembership{Student~member,~IEEE},
        Rong~Zheng,~\IEEEmembership{Senior Member,~IEEE}
\thanks{Manuscript received 16 January 2023; revised 14 April 2023; accepted
	14 August 2023. \textit{(Corresponding Author: Cai Wen; Rong Zheng.)}}
\thanks{Wei Zhao, Rong Zheng and Quan Yuan are with 
	the department of Computing and Software, McMaster University, Hamilton, ON L8S 4L8 Canada (e-mail: zhaow9@mcmaster.ca, rzheng@mcmaster.ca, yuanq4@mcmaster.ca).}
\thanks{Cai Wen is with the School of Information Science and Technology, Northwest University, Xi’an, China. He is also with the department of Computing and Software, McMaster University, Hamilton, Canada. (e-mail: wencai@nwu.edu.cn).}
}

\markboth{Journal of \LaTeX\ Class Files,~Vol.~18, No.~9, September~2020}%
{How to Use the IEEEtran \LaTeX \ Templates}

\maketitle

\begin{abstract}
	Rotating Synthetic Aperture Radar (ROSAR) can generate a 360\degree~image of its surrounding environment using the collected data from a single moving track. Due to its non-linear track, the Back-Projection Algorithm (BPA) is commonly used to generate SAR images in ROSAR. Despite its superior imaging performance, BPA suffers from high computation complexity, restricting its application in real-time systems. In this paper, we propose an efficient imaging method based on robust sparse array synthesis. It first conducts range-dimension matched filtering, followed by azimuth-dimension matched filtering using a selected sparse aperture and filtering weights\blue{. The aperture and weights are computed offline in advance} to ensure robustness to array manifold errors induced by the imperfect radar rotation. We introduce robust constraints on the main-lobe and sidelobe levels of filter design. The resultant robust sparse array synthesis problem is a non-convex  optimization problem with quadratic constraints. An algorithm based on feasible point pursuit and successive convex approximation is devised to solve the optimization problem. Extensive simulation study and experimental evaluations using a real-world hardware platform demonstrate that the proposed algorithm can achieve image quality comparable to that of BPA, but with a substantial reduction in computational time up to 90\%. 
\end{abstract}


\section{Introduction}
Synthetic Aperture Radar (SAR) has been widely used in military reconnaissance and remote sensing, because of its all-weather all-day acquisition capabilities \cite{willey1985synthetic}. Conventional SAR working modes include ``stripmap", ``spotlight" and ``scan"\cite{moreira1996extended}. In these modes, high range resolutions are achieved by transmitting large bandwidth signals, while the high resolution in the cross-range dimension is achieved by utilizing the Doppler effect induced by the relative motion between the radar platform and the target. However, the imaging swaths of these SAR modes are relatively small due to the limited beam footprint and the restricted moving track. Different from the aforementioned imaging schemes, Rotating SAR (ROSAR) systems mount antennas on the edge of rotation platforms with a certain radius\cite{klausing1989feasibility}. Through platform rotating, ROSAR systems are able to scan the surrounding environment continuously and generate a 360\degree~image using the collected data from a single moving track\cite{ali2011short}. ROSAR can overcome limited (angular) field-of-view of radar boards and allow imaging without translational movements of the platform \blue{making it a promising low-cost solution in helicopter-borne SAR imaging\cite{li2015novel, nan2022panoramic, zhang2018wavenumber}, indoor imaging\cite{ali2014rotating} and so on. In indoor environments, ROSAR can be used for mapping and localization in case of fire emergencies or situations where other sensors fail due to high heat and low visibility.}

Due to the highly non-linear moving track of ROSAR, Back-Projection Algorithm (BPA)\cite{cumming1979digital, zeng2006back} is typically employed, where its basic idea is to perform range-azimuth matched filtering with the prior knowledge of the distance between the target and each phase center. Although the conventional BPA can produce high-quality images without any limitation on the array geometry, it suffers extremely high computation complexity making it inadequate for  real-time high-resolution imaging systems. The computational complexity of BPA is proportional to the number of pixels, the number of fast-time samples per pulse and the number of pulses needed to generate one image. In a practical system, all three parameters can be very large: the number of pixels depends on the image resolution; a high pulse repetition frequency (PRF) and consequently dense virtual array elements are required to avoid  aliasing\cite{yanik2019near}; and a large signal bandwidth, which results in a large number of fast-time samples, is needed to ensure the high range-resolution. However, due to the unique array geometry of ROSAR, frequency domain processing algorithms such as Chirp Scaling Algorithm\cite{raney1994precision} and Omega-K\cite{zhu2017range} that assume linear motions of the radar platform relative to the scene are not applicable. In the past decades, much effort has been made in improving the efficiency of BPA \blue{and many algorithms have been proposed}. For example, fast factorized Back-Projection (FFBP)\cite{ulander2003synthetic, zhang2014fast}, Cartesian factorized BPA\cite{dong2017cartesian} and its variant\cite{luo2018modified}. The core idea of these algorithms is sub-aperture fusion. In sub-aperture fusion, the entire aperture is split into many small apertures and BPA is applied to each sub-aperture to obtain coarse images. A high-quality image can then be obtained by fusing these coarse-grained images together. However, \blue{all of them assume a linear aperture too and cannot be applied to circular aperture directly. In addition, sparse array synthesis  is also a technique with low complexity. Conventional ways to select sparse elements, e.g., randomly or uniformly, are not optimal under every condition. Compressive sensing-based algorithms \cite{bi2019theory, zhang2012sparse} still suffer from high complexity, the requirement of sparse environment and sensitivity to array manifold error.}

In this work, we propose a \blue{new} sparse array synthesis technique to reduce the computation complexity of BPA\cite{wen2021reconfigurable}. A key novelty of our design lies in the consideration of array geometry mismatch. Mismatch is prevalent in practical ROSAR systems due to imperfect rotational control or measurements. To solve for sparse complex weights of the virtual array elements, we formulate a robust constrained optimization problem and devise an algorithm based on feasible point pursuit (FPP) \cite{mehanna2014feasible} and successive convex approximation (SCA) \cite{beck2010sequential}. \blue{Compared with conventional aforementioned method, the proposed method is optimal subject to sidelobe constraints and robust to a certain level of array manifold error. Besides, thanks} to the symmetry of the circular array, the algorithm only needs to be executed in an offline manner for one azimuth direction per range bin in the radar coverage area\blue{, and the results can be used for range bins in any direction}. The resulting sparse weights effectively reduce the number of pulses \blue{needed} in BPA. To further reduce the complexity of the proposed algorithm, we perform range-dimension matched filtering by employing Fast Fourier Transform (FFT). For a specific target in space, only the signals from the appropriate range bins at each phase center is selected. \blue{The sparse array design constitutes an important step toward realizing ROSAR on mobile devices with limited in space, battery power and computation capacity.}

We have implemented the proposed algorithms in MATLAB. Extensive numerical simulations are conducted to evaluate the impact of the parameter settings on the sparsity of the design and array patterns. Additionally, we simulate radar transmission and receiving signals using the MATLAB Phased Array Toolbox and collect real-world data from indoor environments from a rotational hardware platform. The evaluation study shows that in both simulations and real experiments the proposed algorithm can reduce the total computation time by more than 90\% while generating SAR images with comparable quality as BPA.

The rest of the paper is organized as follows. Section \ref{sec:SystemModel} gives the system model of ROSAR and formulates the sparse array synthesis problem for ROSAR. Problem transformation and the solution approach are  proposed in Section \ref{sec:SolutionApproach}. Section \ref{sec:Time-domainComputationComplexityReduction} introduces range-dimension filtering using range-FFT to further reduce computation complexity. We validate our approach in Section \ref{sec:SimulationStudy} by numerical evaluation and simulation study as well as experiments in real environment in Section \ref{sec:Testbed Evaluation}. Section \ref{sec:Conclusion} concludes the paper.

\section{System Model and Problem Formulation} \label{sec:SystemModel}
In this section, we introduce the ROSAR system geometry, signal
model, preprocessing steps and give the formal problem
formulation of spare array design at the end. Frequently used symbols are summarized in
Table \ref{tab:notation}.

\begin{table}
	\caption {Frequently Used Symbols} \label{tab:notation} 
	\begin{tabular}{|c|m{6.5cm}|}
		\hline
		\(N\) & \# of virtual phase centers per circle \\
		\hline
		\(n\) & Index of phase centers \\
		\hline
		\(\phi_{n}\) & Bore-sight direction of the \(n\)-th phase center \\
		\hline
		\(\phi_{t}\), \(R_{t}\) & The direction and range of target \(t\) \\
		\hline
		\(\phi_{v}\) & The angle such that if $\phi_n \in (\pi/2 - \phi_v, \pi/2 + \phi_v)$, the target is visible to the $n$-th phase center \\
		\hline
		\(\theta_{n}\), \(R_{n}\) & The distance and direction observed from the \(n\)-th phase center to the target \\
		\hline
		\(p( \cdot )\) & Antenna radiation pattern \\
		\hline
		\(x(t)\) & Transmitted chirp signal \\
		\hline
		\(y_{n}(t)\) & Received signal at the \(n\)-th phase center \\
		\hline
		\(y_{\text{IF},n}(t)\) & IF signal at the \(n\)-th phase center \\
		\hline
		\(y_{\text{IF},n}(m)\) & Sampled IF signal at the \(n\)-th phase center \\
		\hline
		\(y_{\text{1D},n}(l)\) & The data in the \(l\)-th range bin at the \(n\)-th phase center \\
		\hline
		\(\mathbf{y}_{\text{1D},n}\) & Data vector after applying range-FFT at the \(n\)-th phase center \\
		\hline
		\(\mathbf{Y}_{\text{IF}}\), \(\mathbf{Y}_{\text{1D}}\) & Valid data matrix for SAR \\
		\hline
		\(\mathbf{a}(\phi;\ R)\) & Steering vector \\
		\hline
		\(F(\phi;R)\) & Array pattern \\
		\hline
		\(\mathbf{w}\) & Weight vector \\
		\hline
		\(\mathbf{e}\) & Array error vector \\
		\hline
		\(\left( \cdot \right)^{T}\) & Matrix transpose \\
		\hline
		\(\left( \cdot \right)^{H}\) & Matrix conjugate-transpose \\
		\hline
	\end{tabular}
\end{table}

\subsection{Radar Geometry}
\begin{figure}[ht]
	\centering
	\includegraphics[width=0.66\linewidth]{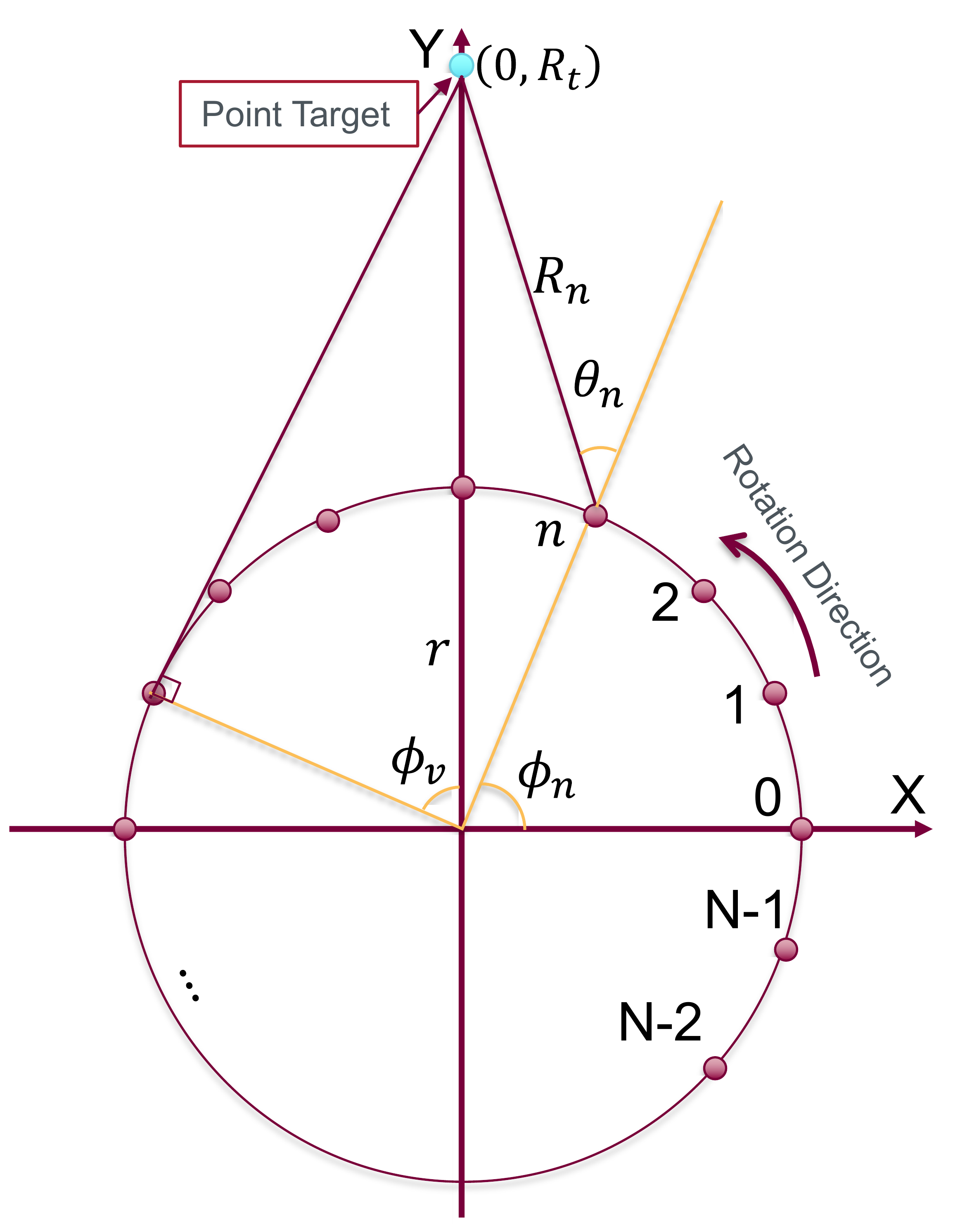}
	\caption{Imaging geometry of a ROSAR system}
	\label{fig:SystemModel}
\end{figure}
Consider a stationary ROSAR system in Fig. \ref{fig:SystemModel}. The radar is moving along the
edge of a circle centered at the origin with radius \(r\). The bore-sight of the
antenna always faces outwards along the radial directions. The antenna
radiation pattern in azimuth is assumed to be cosine-shape and non-zero
within \(\left\lbrack - \frac{\pi}{2},\frac{\pi}{2} \right\rbrack\). The
radar transmits chirp signals at a constant rate, e.g.,
\(N\) times per circle. Due to the symmetry, without loss of generality, we define a 2D coordinate frame such that a point target $R_t$ distance away from the circle center locates at $(0, R_t)$ and the first (indexed by 0) phase center (with respect to the $X$-axis counter-clock wise) is at $(r, 0)$. Then, the bore-sight direction of the antenna at the
\(n\)-th radar position (phase center) is
\begin{equation}
	\phi_{n} = \frac{2\pi n}{N},
\end{equation}
where \(n = 0,1,..,N - 1\). Let \(R_{n}\) and
\(\theta_{n}\) be the distance and the direction from the \(n\)-th
phase center to the target with respect to its bore-sight direction, respectively. Due to the cosine antenna beam pattern, the target is in the field of view (FoV) of only a subset of antenna positions. Denote by $\phi_v$ the angle such that if $\phi_n \in (\pi/2 - \phi_v, \pi/2 + \phi_v)$, the target is visible to the $n$-th phase center.  
\(R_{n}\), \(\theta_{n}\) and \(\phi_{v}\) can be derived from trigonometry relationships, i.e.,
\begin{eqnarray}
	R_{n} &=& \sqrt{R_{t}^{2} + r^{2} - 2rR_{t}\cos\left( \phi_{n} - \phi_{t} \right)},\label{R_n}\\
	\theta_{n} &=& \arctan\frac{R_{t} \cdot \sin\left( \left| \phi_{n} - \phi_{t} \right| \right)}{R_{t} \cdot \cos\left( \left| \phi_{n} - \phi_{t} \right| \right) - r},\\
	\phi_{v} &=& \arccos\frac{r}{R_{t}}.
\end{eqnarray}
The indices of the phase centers where the target is in their FOV
are given by
\begin{equation}
	n = N_{\min},N_{\min} + 1,N_{\min} + 2,\ldots,N_{\max}, \label{n}
\end{equation}
where
\(N_{\min} = \left\lceil \frac{\frac{\pi}{2} - \phi_{v}}{\phi_{\Delta}} \right\rceil\),
\(N_{\max} = \left\lfloor \frac{\frac{\pi}{2} + \phi_{v}}{\phi_{\Delta}} \right\rfloor\)
and \(\phi_{\Delta} = \frac{2\pi}{N}\). To generate the image of the target, only
signals received by those phase centers are used.

\subsection{Signal Model and Preprocessing}
Let the chirp signal transmitted by the radar be
\begin{equation}
	x(t) = e^{j2\pi\left( f_{c}t + \frac{1}{2}Kt^{2} \right)},
\end{equation}
where \(f_{c}\) is the carrier frequency, \(K\) is the chirp slope and
\(t\) is the fast time. The received signal at the \(n\)-th phase center
is
\begin{equation}
	y_{n}(t) = \alpha_{n}e^{j2\pi\left\lbrack f_{c}(t - \tau_{n}) + \frac{1}{2}K(t - \tau_{n})^{2} \right\rbrack} + v(t),
\end{equation}
where \(\alpha_{n}\) \blue{combines the complex reflection coefficient
	of the target, the antenna radiation pattern and channel fading}, \(\tau_{n} = {2R_{n}}/{c}\) is
the round-trip time delay, \(c\)
is the speed of the light, \(v(t)\) is the Gaussian white noise at the receiver
side. Specifically, the antenna radiation pattern is represented as
\begin{equation}
	p\left( \theta \right) = \left\{ \begin{matrix}
		\cos\left( \theta \right) & \theta \in \left( - \frac{\pi}{2},\frac{\pi}{2} \right), \\
		0 & \text{otherwise}. \\
	\end{matrix} \right.
\end{equation}
After down-converting and deramping, the resulting
intermediate frequency (IF) signal is
\begin{equation}
	\begin{array}{r@{}l}
		y_{\text{IF},n}(t) =& y_{n}(t)^{*} \cdot x(t)\\
		=& \alpha_{n}e^{j2\pi\left\lbrack \tau_{n} Kt + \left( f_{c}\tau_{n} - \frac{1}{2}K\tau_{n}^{2} \right) \right\rbrack} + v_{\text{IF}}(t)\\
		\approx& \alpha_{n}e^{j2\pi\left\lbrack \tau_{n} Kt + f_{c}\tau_{n} \right\rbrack} + v_{\text{IF}}(t),
	\end{array}
\end{equation}
where \(v_{\text{IF}}(t) = v^{*}(t) \cdot x(t)\). The residual video phase
(RVP) term \(- \frac{1}{2}K\tau_{n}^{2}\) is negligible compared with
other phase terms. Let the sampling frequency, the total number of samples, sampling
interval, sampling start time in one chirp be \(F_{s}\), \(M\), \(t_{s}\)
(\(t_{s} = {1}/{F_{S}}\)), \(T_{Start}\), respectively. The sampled
IF signal is
\begin{equation}
	y_{\text{IF},n}(m) = \alpha_{n}e^{j2\pi\left\lbrack \tau_{n} K\left( mt_{s} + T_{Start} \right) + f_{c}\tau_{n} \right\rbrack} + v_{\text{IF}}(m),\label{y_IF_n_m}
\end{equation}
where \(m\) is the sampling index, \(0 \leq m \leq M - 1\). Combining
all the samples, we get the following vector representation
\begin{equation}
	\mathbf{y}_{\text{IF},n} = \begin{bmatrix}
		\alpha_{n}e^{j2\pi\left\lbrack \tau_{n} K\left( 0 \cdot t_{s} + T_{Start} \right) + f_{c}\tau_{n} \right\rbrack} \\
		\alpha_{n}e^{j2\pi\left\lbrack \tau_{n} K\left( 1 \cdot t_{s} + T_{Start} \right) + f_{c}\tau_{n} \right\rbrack} \\
		\vdots \\
		\alpha_{n}e^{j2\pi\left\lbrack \tau_{n} K\left( (M - 1) \cdot t_{s} + T_{Start} \right) + f_{c}\tau_{n} \right\rbrack} \\
	\end{bmatrix} + \mathbf{v}_{\text{IF}},\label{y_n_IF}
\end{equation}
where
\(\mathbf{v}_{\text{IF}} = \left\lbrack v_{\text{IF}}(0),v_{\text{IF}}(1),\ldots,v_{\text{IF}}(M - 1) \right\rbrack^{T}\). Let \(k = {2\pi\left( KT_{start} + f_{c} \right)}/{c}\). Substituting the \(\tau_{n}\)'s in each entry by \({2R_{n}}/{c}\) and rearranging
items in \eqref{y_n_IF}, we have
\begin{equation}
	\mathbf{y}_{\text{IF},n} = \begin{bmatrix}
		\alpha_{n}e^{j2\pi\tau_{n}K \cdot 0 \cdot t_{s}}e^{j2kR_{n}} \\
		\alpha_{n}e^{j2\pi\tau_{n}K \cdot 1 \cdot t_{s}}e^{j2kR_{n}} \\
		\vdots \\
		\alpha_{n}e^{j2\pi\tau_{n}K \cdot (M - 1) \cdot t_{s}}e^{j2kR_{n}} \\
	\end{bmatrix} + \mathbf{v}_{\text{IF}}.\label{y_n_IF2}
\end{equation}
Now, the data matrix from the effective phase centers used for SAR is
\begin{equation}
	\mathbf{Y}_{\text{IF}} = \left\lbrack \mathbf{y}_{\text{IF},N_{\min}},\mathbf{y}_{\text{IF},N_{\min} + 1},\ldots,\mathbf{y}_{\text{IF},N_{\max}} \right\rbrack.
\end{equation}

The conventional BPA images a point with parameter \(\left( \phi_{t},R_{t} \right)\) by
computing the Hadamard product of \(\mathbf{Y}_{\text{IF}}\) and a matrix
\(\mathbf{W}_{\text{BP}}\) of size
\(M \times \left( N_{\max} - N_{\min} + 1 \right)\), where the element
locating at the \(i\)-th row and \(j\)-th column is
\begin{equation}
	\mathbf{W}_{\text{BP},(i,j)} \!=\! \alpha_{N_{\min} + j - 1}e^{-j2\pi\tau_{N_{\min} + j - 1}K(i - 1)t_{s}}e^{-j2kR_{N_{\min} + j - 1}}. \nonumber
\end{equation}
Then, the intensity of the point is
\begin{equation}
	I\left( \phi_{t},R_{t} \right) = \mathbf{1}^{T} \cdot \left( \mathbf{W}_{\text{BP}}\odot\mathbf{Y}_{\text{IF}} \right) \cdot \mathbf{1}, \label{BPA}
\end{equation}
where \(\odot\) is the symbol of Hadamard product. If the target indeed
locates at \(\left( \phi_{t},R_{t} \right)\) in polar coordinates, all the phases of the
sampled data are perfectly compensated, and \eqref{BPA} achieves its maximum. However, the computation complexity of imaging a rectangular area using
conventional BPA is \(O\left( L_{x} \times L_{y} \times M \times N \right)\), where \(L_{x}\)
and \(L_{y}\) is the number of grids along \(X\) and \(Y\) direction of
the area. Clearly, the complexity grows linearly with \(M\), $N$ and area size. From the
complexity analysis, it can be deduced that two possible ways to lower the complexity of BPA are, (1) reducing the number of phase centers to be used, i.e., reducing \(N\), and (2) apply range-dimension matched filtering and select the appropriate range bin instead of using all \(M\) data samples in each pulse.

In the subsequent sections, we first develop a Sparse Array Synthesis (SAS) method that selects a subset of the phase centers and assigns appropriate complex weights. Then, we investigate the use of range-dimension matched filter (or more commonly known as range FFT) to further reduce computation complexity. The two approached are abbreviated to ``SAS'' and ``FFT+SAS'' respectively for simplicity.

\subsection{Problem Formulation for Robust Sparse Array Synthesis}
For simplicity, we assume that the reflection coefficient is always 1. \blue{Due to complex multipath reflection, wall penetration in indoor environments and the small diameter of the rotation platform relative to the dimension of the environment, the channel fading factor can be approximated to be a constant for the same range bin in all directions and thus we omit it in the formulation}. Thus,
\begin{equation}
	\alpha_{n} = p\left( \theta_{n} \right).\label{alpha_n}
\end{equation}
BPA can be viewed as a form of range-azimuth two-dimension filtering. To generalize it to sparsely-selected phase centers, we first apply a compensation matrix to \(\mathbf{Y}_{\text{IF}}\) to remove the phase items related to fast-time sampling, i.e.,
\begin{eqnarray*}
	\mathbf{Y}_{\text{IF}}' &=& \mathbf{W}_{\text{SA}}\odot\mathbf{Y}_{\text{IF}}, \\
	&=& \begin{bmatrix}
		\alpha_{N_{\min}}e^{j2kR_{N_{\min}}} & \cdots & \alpha_{N_{\max}}e^{j2kR_{N_{\max}}} \\
		\alpha_{N_{\min}}e^{j2kR_{N_{\min}}} & \cdots & \alpha_{N_{\max}}e^{j2kR_{N_{\max}}} \\
		\vdots & \ddots & \vdots \\
		\alpha_{N_{\min}}e^{j2kR_{N_{\min}}} & \cdots & \alpha_{N_{\max}}e^{j2kR_{N_{\max}}} \\
	\end{bmatrix},
\end{eqnarray*}
where the \(i\)-th row and \(j\)-th column element of \(\mathbf{W}_{\text{SA}}\) is
\(\mathbf{W}_{\text{SA},(i,j)} = e^{- j2\pi\tau_{N_{\min} + j - 1}K \cdot (i - 1) \cdot t_{s}}\). The steering vector of the ROSAR array to a near-field target located at range \(R\) can be represented as
\begin{equation}
	\mathbf{a}(\phi;R) = \begin{bmatrix}
		\cos( \theta^{'}_{N_{\min}} )  e^{j2k\sqrt{R^{2} + r^{2} - 2Rr\cos( \phi - \phi_{N_{\min}} )}} \\
		\cos( \theta^{'}_{N_{\min}+1} )e^{j2k\sqrt{R^{2} + r^{2} - 2Rr\cos( \phi - \phi_{N_{\min} + 1} )}} \\
		\vdots \\
		\cos( \theta^{'}_{N_{\max}} )e^{j2k\sqrt{R^{2} + r^{2} - 2Rr\cos( \phi - \phi_{N_{\max}} )}} \\
	\end{bmatrix}, \label{OriginalArrayPattern}
\end{equation}
where \( \theta^{'}_{n} = \arctan\frac{R \sin( \left| \phi - \phi_{n} \right| )}{R \cos( \left| \phi - \phi_{n} \right| ) - r} \), and the array pattern can also be calculated as
\begin{equation}
	F(\phi;R) = \mathbf{w}^{H}\mathbf{a}(\phi;R),
\end{equation}
where \(\mathbf{w}\) is a sparse complex weight vector to be designed and some of its elements can be equal or be close to zero. Note that, \(\mathbf{w}\) is also a function of \(R\), but we omit the subscript \(R\) for simplicity. To focus a point target locating at \((\phi_{t}, R_{t})\), we need to
compute
\begin{eqnarray*}
	I\left( \phi_{t},R_{t} \right) \!\!\!&=&\!\!\! \mathbf{1}^{T} \cdot \left( \mathbf{w}^{H} \circ \mathbf{Y}_{\text{IF}}' \right) \cdot \mathbf{1} \\
	\!\!\!&=&\!\!\! \mathbf{1}^{T} \cdot \left( \mathbf{w}^{H} \circ \mathbf{W}_{\text{SA}}\odot\mathbf{Y}_{\text{IF}} \right) \cdot \mathbf{1},
\end{eqnarray*}
where \(\circ\) is the Khatri-Rao product.

Due to the vibration of the rotation platform, odometry errors and antenna pattern mismatch (e.g., cosine pattern), there  exist array manifold errors, which may lead to blurred images. To obtain an SAR image with good quality in this situation, the sparse weight vector \(\mathbf{w}\) must be carefully designed with consideration of robustness to array errors.

We formulate the way to obtain the desirable \(\mathbf{w}\) as to solve an optimization problem. In the optimization, the first constraint is that the power of the main-lobe peak, locating at \(\phi_{m} = \frac{\pi}{2}\), should be larger than or equal to a threshold \(U\),
i.e.,
\begin{equation}
	\left| \mathbf{w}^{H}\left( \mathbf{a}\left( \phi_{m};R \right) + \mathbf{e}_{m} \right) \right|^{2} \geq U,\ \ \left\| \mathbf{e}_{m} \right\| \leq \Delta_{R},
\end{equation}
where \(\mathbf{e}_{m}\) is the array error vector caused by measurement
and imperfect radar rotations, and we assume that \(\mathbf{e}_{m}\) is bounded by a ball with radius \(\Delta_{R}\)
(\(0 \le \Delta_{R} \le \left\| \mathbf{a}(\phi;R) \right\|\)). Second, to
restrict the main-lobe width and the sidelobe level, we put limitations on the received power
at some uniformly spaced discrete directions except for the desired main lobe
area, i.e.,
\begin{equation}
	\left| \mathbf{w}^{H}\left( \mathbf{a}\left( \phi_{s};R \right) + \mathbf{e}_{s} \right) \right|^{2} \leq \eta U,\ \ s = 1,2,\ldots,S;\left\| \mathbf{e}_{s} \right\| \leq \Delta_{R},
\end{equation}
where \(\mathbf{e}_{s}\) is the error vector for sidelobe area with the same properties as \(\mathbf{e}_{m}\), \(S\) is the number of uniformly spaced discrete directions,
\(\phi_{s} \in \left\lbrack \phi_{N_{\min}},\frac{\pi}{2} - \phi_{\text{MW}} \right\rbrack \cup \left\lbrack \frac{\pi}{2} + \phi_{\text{MW}},\phi_{N_{\max}} \right\rbrack\),
\(\phi_{\text{MW}}\) is the half of the desirable main-lobe width, \(\eta\) is the pre-set power ratio of the main-lobe to the sidelobe. Third, to avoid amplifying the noise level, we impose a constraint on the gain of noise power:
\begin{equation}
	\left\| \mathbf{w} \right\|_{2}^{2} = 1.
\end{equation}
Lastly, to guarantee a sufficient gain on the target, we set another constraint \(U \geq U_{\min}\). 

The objective is to minimize the number of
virtual phase centers given by \( \left\| \mathbf{w} \right\|_{0}\). To this end, we formulate the sparse array synthesis problem as
\begin{equation}
	\min_{\mathbf{w},U,\mathbf{e}_{m},\mathbf{e}_{s}}\left\| \mathbf{w} \right\|_{0}\nonumber
\end{equation}
\begin{equation}
	s.t.\!\!\left\{ \!\!\!\!\begin{array}{l}
		C1\!\!:\!\!\left| \mathbf{w}^{H}\left( \mathbf{a}\left( \phi_{m};R \right) + \mathbf{e}_{m} \right) \right|^{2} \geq U,\left\| \mathbf{e}_{m} \right\| \leq \Delta_{R}, \\
		C2\!\!:\!\!\left| \mathbf{w}^{H}( \mathbf{a}( \phi_{s};R ) \!+\! \mathbf{e}_{s} ) \right|^{2} \!\!\leq\!\! \eta U,s \!=\! 1,2,\ldots,S;\!\left\| \mathbf{e}_{s} \right\| \!\!\leq\!\! \Delta_{R}, \\
		C3\!\!:\!\!\left\| \mathbf{w} \right\|_{2}^{2} = 1, \\
		C4\!\!:\!\!U \geq U_{\min}. \\
	\end{array} \right.
	\label{OriginalOptimizationProblem}
\end{equation}

Problem \eqref{OriginalOptimizationProblem} is a non-convex optimization problem since both the objective function and constraints C1 and C3 are non-convex. Because of the consideration of robustness to array errors, the problem formulation is markedly different from those in conventional sparse array synthesis\cite{nai2010beampattern,fuchs2012synthesis,liu2008reducing,liu2009reducing,keizer2008linear,du2011weighted}, rendering existing techniques inapplicable. In the next section, we develop a customized algorithm based on FPP and SCA to solve \eqref{OriginalOptimizationProblem}.

\section{Solution Approach for Robust Sparse Array Synthesis (SAS)} \label{sec:SolutionApproach}
\subsection{Problem Transformation}
Directly solving \eqref{OriginalOptimizationProblem} is hard, since \(l^{0}\)-norm minimization problem requires intractable combinatorial search. To reduce the complexity, we replace the \(l^{0}\)-norm objective
function with \(l^{1}\)-norm, i.e., \( \left\| \mathbf{w} \right\|_{1} \) as suggested by \cite{candes2008enhancing, donoho2006compressed}.

C1 and C2 contain additional control variables $\mathbf{e}_m$ and $\mathbf{e}_s$ to express robustness constraints. They can be simplified by considering the worst case scenarios. Specifically, by using the Cauchy-Schwarz inequality and the triangle inequality, we can find the minimum of main-lobe response and the maximum of sidelobe response respectively as follows
\begin{eqnarray}
	\begin{array}{r@{}l}
		\left| \mathbf{w}^{H}( \mathbf{a}( \phi_{m};R ) + \mathbf{e}_{m} ) \right|^{2}\!\!=& \left| \mathbf{w}^{H}\mathbf{a}( \phi_{m};R ) + \mathbf{w}^{H}\mathbf{e}_{m} \right|^{2}\\
		\geq& ( | \mathbf{w}^{H}\mathbf{a}( \phi_{m};R ) | - \left| \mathbf{w}^{H}\mathbf{e}_{m} \right| )^{2}\\
		\geq& ( | \mathbf{w}^{H}\mathbf{a}( \phi_{m};R ) | - \left\| \mathbf{w} \right\|_{2}\left\| \mathbf{e}_{m} \right\|_{2} )^{2}\\
		\geq& ( | \mathbf{w}^{H}\mathbf{a}( \phi_{m};R ) | - 1 \cdot \Delta_{R} )^{2},
	\end{array}
	\label{C1Relaxation}
\end{eqnarray}
\begin{eqnarray}
	\begin{array}{r@{}l}
		\left| \mathbf{w}^{H}( \mathbf{a}( \phi_{s};R ) + \mathbf{e}_{s} ) \right|^{2} \!\!=& \left| \mathbf{w}^{H}\mathbf{a}( \phi_{s};R ) + \mathbf{w}^{H}\mathbf{e}_{s} \right|^{2}\\
		\leq& ( | \mathbf{w}^{H}\mathbf{a}( \phi_{s};R ) | + \left| \mathbf{w}^{H}\mathbf{e}_{s} \right| )^{2}\\
		\leq& ( | \mathbf{w}^{H}\mathbf{a}( \phi_{s};R ) | + \left\| \mathbf{w} \right\|_{2}\left\| \mathbf{e}_{s} \right\|_{2} )^{2}\\
		\leq& ( | \mathbf{w}^{H}\mathbf{a}( \phi_{s};R ) | + 1 \cdot \Delta_{R} )^{2}.
	\end{array}
	\label{C2Relaxation}
\end{eqnarray}
The equalities hold when \(\mathbf{e} = a\cdot\mathbf{a}( \phi;R )\) (\(a\in \mathbb{R}\)). Substituting \eqref{C1Relaxation} and \eqref{C2Relaxation} into C1 and C2, we obtain the worst-case constraints as
\begin{eqnarray}
	( | \mathbf{w}^{H}\mathbf{a}( \phi_{m};R ) | - \Delta_{R} )^{2} \!\!&\geq&\!\! U \label{C1RelaxationResult}\\
	( | \mathbf{w}^{H}\mathbf{a}( \phi_{s};R ) | + \Delta_{R} )^{2} \!\!&\leq&\!\!  \eta U,s = 1,2,\ldots,S \label{C2RelaxationResult}
\end{eqnarray}
Taking the square root of both sides of \eqref{C1RelaxationResult} and \eqref{C2RelaxationResult}, and re-arranging
items in the inequalities, we have
\begin{equation}
	C1:( U^{'} + \Delta_{R} )^{2} - \mathbf{w}^{H}\mathbf{a}( \phi_{m};R )\mathbf{a}^{H}( \phi_{m};R )\mathbf{w} \leq 0,\label{C1squareroot}
\end{equation}
\begin{eqnarray}
	C2:\mathbf{w}^{H}\mathbf{a}( \phi_{s};R )\mathbf{a}^{H}( \phi_{s};R )\mathbf{w} - ( \sqrt{\eta}U^{'} - \Delta_{R} )^{2} \leq 0,\nonumber \\
	s = 1,2,\ldots,S,\label{C2squareroot}
\end{eqnarray}
where \(U^{'} = \sqrt{U}\). \(C3\) in \eqref{OriginalOptimizationProblem} can be replaced by two
inequality constraints as
\begin{equation}
	C3:\left\| \mathbf{w} \right\|_{2}^{2} - 1 \leq 0,
\end{equation}
\begin{equation}
	C4:1 - \left\| \mathbf{w} \right\|_{2}^{2} \leq 0,
\end{equation}
and \(C4\) is replaced by
\begin{equation}
	C5:U^{'} \geq U_{\min}^{'},
\end{equation}
where \(U_{\min}^{'} = \sqrt{U_{\min}}\). The optimization problem is thus transformed to
\begin{equation}
	\min_{\mathbf{w},U^{'}}\left\| \mathbf{w} \right\|_{1}\nonumber
\end{equation}
\begin{equation}
	s.t.\!\!\left\{ \!\!\!\!\begin{array}{l}
		C1\!:\!( U^{'} + \Delta_{R} )^{2} - \mathbf{w}^{H}\mathbf{a}( \phi_{m};R )\mathbf{a}^{H}( \phi_{m};R )\mathbf{w} \leq 0, \\
		C2\!:\!\mathbf{w}^{H}\mathbf{a}( \phi_{s};R )\mathbf{a}^{H}( \phi_{s};R )\mathbf{w} - ( \sqrt{\eta}U^{'} - \Delta_{R} )^{2} \leq 0, \\
		\hfill s = 1,2,\ldots,S, \\
		C3\!:\!\left\| \mathbf{w} \right\|_{2}^{2} - 1 \leq 0, \\
		C4\!:\!1 - \left\| \mathbf{w} \right\|_{2}^{2} \leq 0, \\
		C5\!:\!U^{'} \geq U_{\min}^{'}. \\
	\end{array} \right.\!\!\!\!
	\label{ReformulatedOptimizationProblem}
\end{equation}

\subsection{FPP-SCA-based Algorithm}
The reformulated problem \eqref{ReformulatedOptimizationProblem} is still non-convex which is hard to be solved directly, and it is also tricky to find feasible initial solutions, but it is now amenable to the convex approximation technique. Inspired by the idea of FPP \cite{mehanna2014feasible}, we introduce three slack variables \(b\),
\(b_{1}\), \(b_{2}\) (\(b,b_{1},b_{2} > 0\)) for C1--C4, and we construct the following slacked surrogate problem of \eqref{ReformulatedOptimizationProblem}
\begin{equation}
	\min_{\mathbf{w},U^{'},b,b_{1},b_{2}}{\left\| \mathbf{w} \right\|_{1} + \lambda_{b}\left( b + b_{1} + b_{2} \right)}\nonumber
\end{equation}
\begin{equation}
	s.t.\!\!\left\{ \!\!\!\!\begin{array}{l}
		C1\!:\!( U^{'} + \Delta_{R} )^{2} - \mathbf{w}^{H}\mathbf{a}( \phi_{m};R )\mathbf{a}^{H}( \phi_{m};R )\mathbf{w -}b_{1} \leq 0, \\
		C2\!\!:\!\! \mathbf{w}^{H}\mathbf{a}( \phi_{s};R )\mathbf{a}^{H}( \phi_{s};R )\mathbf{w} - ( \sqrt{\eta}U^{'} - \Delta_{R} )^{2} - b_{2} \leq 0, \\
		\hfill s = 1,2,\ldots,S, \\
		C3\!:\!\left\| \mathbf{w} \right\|_{2}^{2} - 1 - b \leq 0, \\
		C4\!:\!1 - b - \left\| \mathbf{w} \right\|_{2}^{2} \leq 0, \\
		C5\!:\!U^{'} \geq U_{\min}^{'},
	\end{array} \right.
	\label{SurrogateOptimizationProblem}
\end{equation}
where \(\lambda_{b}\) (\(\lambda_{b} > 0\)) is a pre-set coefficient to penalize constraint violations.

To handle the non-convex terms in above constraints, i.e.,
\(- \mathbf{w}^{H}\mathbf{a}\left( \phi_{m};R \right)\mathbf{a}^{H}\left( \phi_{m};R \right)\mathbf{w}\)
in \(C1\), \(- ( \sqrt{\eta}U^{'} - \Delta_{R} )^{2}\) in
\(C2\) and \(- \left\| \mathbf{w} \right\|_{2}^{2}\) in \(C4\), we
next apply SCA \cite{beck2010sequential}. SCA is an iterative method and its core idea is to replace non-convex terms with convex approximations (usually upper bounds of these non-convex terms) in each iteration. In our case, given \(\{ \mathbf{w}_{(i)},U_{(i)}^{'} \}\) after the \(i\)-th iteration, the convex upper bounds of the three non-convex terms are derived by applying their respective first-order Taylor expansions as

\begin{equation}
	\begin{array}{r@{}l}
		&- \mathbf{w}^{H}\mathbf{a}\left( \phi_{m};R \right)\mathbf{a}^{H}\left( \phi_{m};R \right)\mathbf{w}\\
		\le& - \mathbf{w}_{(i)}^{H}\mathbf{a}\left( \phi_{m};R \right)\mathbf{a}^{H}\left( \phi_{m};R \right)\mathbf{w}_{(i)}\\
		&- 2\text{Re}\{ \mathbf{w}_{(i)}^{H}\mathbf{a}\left( \phi_{m};R \right)\mathbf{a}^{H}\left( \phi_{m};R \right)\left( \mathbf{w} - \mathbf{w}_{(i)} \right) \}\\
		=& \mathbf{w}_{(i)}^{H}\mathbf{a}\left( \phi_{m};R \right)\mathbf{a}^{H}\left( \phi_{m};R \right)\mathbf{w}_{(i)}\\
		&- 2\text{Re}\{ \mathbf{w}_{(i)}^{H}\mathbf{a}\left( \phi_{m};R \right)\mathbf{a}^{H}\left( \phi_{m};R \right)\mathbf{w} \},
	\end{array}\label{DerivationOfC1NonConvex}
\end{equation}
\begin{equation}
	\begin{array}{r@{}l}
		&- ( \sqrt{\eta}U^{'} - \Delta_{R} )^{2}\\
		\le&- ( \sqrt{\eta}U_{(i)}^{'} - \Delta_{R} )^{2} - 2( \eta U_{(i)}^{'} - \sqrt{\eta}\Delta_{R} )( U^{'} - U_{(i)}^{'} )\\
		=&2( \sqrt{\eta}\Delta_{R} - \eta U_{(i)}^{'} )U^{'} + \eta{U_{(i)}^{'2}} - \Delta_{R}^{2},
	\end{array}\label{DerivationOfC2NonConvex}
\end{equation}
\begin{eqnarray}
	\begin{array}{r@{}l}
		- \left\| \mathbf{w} \right\|_{2}^{2} &\le - \left\| \mathbf{w}_{(i)} \right\|_{2}^{2} - 2Re\{ \mathbf{w}_{(i)}^{H}\left( \mathbf{w} - \mathbf{w}_{(i)} \right) \}\\
		&= \left\| \mathbf{w}_{(i)} \right\|_{2}^{2} - 2Re\{ \mathbf{w}_{(i)}^{H}\mathbf{w} \}.
	\end{array}\label{DerivationOfC4NonConvex}
\end{eqnarray}

Replacing the non-convex terms in C1, C2 and C4 with the convex upper bounds in \eqref{DerivationOfC1NonConvex}--\eqref{DerivationOfC4NonConvex},  we finally transform \eqref{OriginalOptimizationProblem} to a convex sub-problem in \eqref{FinalOptimizationProblem}.
\begin{strip}
	\rule{\linewidth}{1pt}
	\begin{equation}
		\min_{\mathbf{w},U^{'},b,b_{1},b_{2}}{\left\| \mathbf{w} \right\|_{1} + \lambda_{b}\left( b + b_{1} + b_{2} \right)}\nonumber
	\end{equation}
	\begin{equation}
		s.t.\ \left\{ \begin{array}{l}
			C1:( U^{'} + \Delta_{R} )^{2} + \mathbf{w}_{(i)}^{H}\mathbf{a}\left( \phi_{m};R \right)\mathbf{a}^{H}\left( \phi_{m};R \right)\mathbf{w}_{(i)} - 2\text{Re}\{ \mathbf{w}_{(i)}^{H}\mathbf{a}\left( \phi_{m};R \right)\mathbf{a}^{H}\left( \phi_{m};R \right)\mathbf{w} \} - b_{1} \leq 0, \\
			C2:\mathbf{w}^{H}\mathbf{a}\left( \phi_{s};R \right)\mathbf{a}^{H}\left( \phi_{s};R \right)\mathbf{w} + 2( \sqrt{\eta}\Delta_{R} - \eta U_{(i)}^{'} )U^{'} + \eta{U_{(i)}^{'}}^{2} - \Delta_{R}^{2} - b_{2} \leq 0,\ \ s = 1,2,\ldots,S, \\
			C3:\left\| \mathbf{w} \right\|_{2}^{2} - 1 - b \leq 0, \\
			C4:1 - b + \left\| \mathbf{w}_{(i)} \right\|_{2}^{2} - 2\text{Re}\{ \mathbf{w}_{(i)}^{H}\mathbf{w} \} \leq 0, \\
			C5:U^{'} \geq U_{\min}^{'}. \\
		\end{array} \right. \label{FinalOptimizationProblem}
	\end{equation}
	\rule{\linewidth}{1pt}
\end{strip}

Repeatedly solving \eqref{FinalOptimizationProblem} with the values from the previous iteration until \blue{the number of iterations reaches a pre-set value \(ITER\)}. Let
\(J^{(i)} = \left\| \mathbf{(w)}_{i}^{\star} \right\|_{1} + \lambda_{b}( b^{(i)} + b_{1}^{(i)} + b_{2}^{(i)} )\)
be the value of the objective function after the \(i\)-th iteration, \(\mathbf{R}\) be a vector of target range bins of concern, \({\Delta}_{\mathbf{R}}\) be a vector of the maximum \(l^2\)-norm of \(\mathbf{e}\) for all range bins.  Algorithm \ref{algo:Pre-Compute} and \ref{algo:ROSAR_SA} summarize the proposed approaches to determine the sparse weight vector for each range bin and to apply the resulting weight vector for SAR imaging, respectively.

\begin{algorithm}
	\caption{Computing sparse weight vectors}
	\begin{algorithmic}
		\State Input \(\mathbf{R}\), \(\Delta_{\mathbf{R}}\), \(\phi_{MW}\), \(\eta\), \(\lambda_{b}\), \(N\), \(J\), \(\mathbf{w}_{(0)}\), \(U_{(0)}^{'}\);
		\State \({W}^{\star} \leftarrow \varnothing\); \(i \leftarrow 0\);
		\BeginForeach{\(R\), \(\Delta_{R}\)}{\(\mathbf{R}\), \(\Delta_{\mathbf{R}}\)}
		\State Calculate \(\phi_{v}\), \(\phi_{\Delta}\), \(N_{\min}\), \(N_{\max}\), \(a\left( \phi_{m};R \right)\), \(a\left( \phi_{s};R \right)\)
		\Repeat
		\State \(i \leftarrow i + 1\);
		\State Compute \(\mathbf{w}_{(i)}\), \(U_{(i)}^{'}\) by solving \eqref{FinalOptimizationProblem} with	\(\mathbf{w}_{(i - 1)}\), \(U_{(i - 1)}^{'}\);
		\Until{\blue{\( i > ITER \)}}
		\State Save \(\mathbf{w}_{(i)}\) to the set \({W}^{\star}\);
		\EndForeach
		\State Output \({W}^{\star}\).
	\end{algorithmic}
	\label{algo:Pre-Compute}
\end{algorithm}
\begin{algorithm}
	\caption{The proposed approach for ROSAR imaging}
	\begin{algorithmic}
		\State Input \(\mathbf{Y}_{\text{IF}}\), \(R_{t}\), \(\phi_{t}\);
		\State Compute the index of range bin, i.e., \(\mathcal{I}_{R_{t}}\), for \(R_{t}\);
		\State Compute the central index of the phase center \(\mathcal{I}_{\phi_{t}} \leftarrow round(\phi_{t}/\phi_{\Delta})\);
		\State \(\mathbf{w}^{\star} \leftarrow {W}^{\star}\left( \mathcal{I}_{R_{t}} \right)\)\footnotemark;
		\State \(\mathbf{Y}_{\text{IF}}^{'} \leftarrow \mathbf{Y}_{\text{IF}}( :,\mathcal{I}_{\phi_{t}} - \left( |\mathbf{w}^{\star}| - 1 \right)/{2}:\mathcal{I}_{\phi_{t}} + \left( |\mathbf{w}^{\star}| - 1 \right)/{2} )\);
		\State Output \(I( \phi_{t},R_{t} ) \leftarrow \mathbf{1}^{T} \cdot ( \mathbf{w}^{\star H} \circ ( \mathbf{W}_{\text{SA}}\odot\mathbf{Y}_{\text{IF}}^{'} ) ) \cdot \mathbf{1}\).
	\end{algorithmic}
	\label{algo:ROSAR_SA}
\end{algorithm}

\subsection{Initial values and parameter settings for Algorithms \ref{algo:Pre-Compute} and \ref{algo:ROSAR_SA}}
\subsubsection{Initial Value Settings}\label{InitialValueSelection}
The proposed approach is able to work with any initial values of the control variables since the constraints are always feasible due to the introduced slack variables. In our implementation, the initial value of \(\mathbf{w}\) is chosen as
\(\mathbf{w}_{(0)} = \frac{1}{\sqrt{N_{\max} - N_{\min} + 1}}\mathbf{1}\) and \(U_{(0)}^{'} = | \mathbf{w}_{(0)}^{H}\mathbf{a}(\phi;R) |^{2}\).

\subsubsection{Parameter Settings}\label{ParameterSettings}
\blue{To ensure convergence, \(ITER\) and \(Th\) can be set to around \(50\sim100\) and \(10^{-4}\sim10^{-3}\), respectively. \(\phi_{\text{MW}}\) and \(\eta\) are design parameters of the sparse array. We found that any value lower than both \(\phi_{\text{MW}}=1\degree\) and \(\eta=-33dB\) makes the SCA diverge.} \(\Delta_{R}\) should be chosen by considering practical limitations of target platforms. For example, in Section \ref{sec:HardwareandSoftwareImplementation}, the simulation setup and experiments take the  unstable rotation speed of a ROSAR platform into account.
The direction of each phase center under unstable rotations is modeled as a Gaussian distribution
\begin{equation}
	{\widehat{\phi}}_{n}\sim\mathcal{N}\left( \frac{2\pi n}{N},\sigma \right),\label{phi_n_hat}
\end{equation}
where the \(\sigma\) is the standard deviation of the direction. The error vector is computed from
\(\mathbf{e} = \widehat{\mathbf{a}}(\phi;\ R) - \mathbf{a}(\phi;\ R)\), where
\(\widehat{\mathbf{a}}(\phi;R)\) is determined by substituting
\(\phi_{n}\) with \({\widehat{\phi}}_{n}\). Let
\({\widehat{\Delta}}_{R} = \left\| \mathbf{e} \right\|\). By repeatedly sampling from \eqref{phi_n_hat}, we can obtain the
cumulative distribution function (CDF) of \({\widehat{\Delta}}_{R}\), and choose 
the \({\widehat{\Delta}}_{R}\) corresponding to 99\% of the cumulative
probability as \(\Delta_{R}\). The testbed evaluations
in Section \ref{sec:Testbed Evaluation} show that this model is reasonable in realistic settings. Note since the phase error \(\mathbf{e}\) differs among range bins, one specific \(\Delta_{R}\) must be pre-computed for each range bin.

\blue{To ensure accurate results, the angle interval should be less than or equal to the angular resolution of ROSAR. However, reducing the angle interval increases the number of grid points and leads to higher computation costs. Since there is no closed-form solution to the angular resolution of a circular array, we use the results of a linear array as a reference. The determination of \(U_{\min}\) is based on the expected image quality in target applications. It should be large enough to guarantee a sufficient gain for all range bins. Otherwise, there could be light and dark strips on the generated SAR image.}

\blue{Increasing \(\lambda_{b}\) improves the sparsity of \(\mathbf{w}\) but when \(\lambda_{b}\) is too large, (36) is no longer feasible. As a general rule of thumb, $\lambda_b$ should be several times larger than the maximum of \(\left\| \mathbf{w} \right\|_{1}\) to ensure that the slack variables converge to zero.} Since the \(l^{1}\)-norm differs \(l^{0}\)-norm and cannot enforce sparsity in itself, we must manually set any term in \(\mathbf{w}^{\star}\) lower than a pre-defined threshold to 0. Thus, in Algorithm \ref{algo:ROSAR_SA}, only the nonzero entries in \(\mathbf{w}^{\star}\) are included in computing \(\mathbf{w}^{\star H} \circ \mathbf{W}_{\text{SA}}\odot\mathbf{Y}_{\text{IF}}^{'}\). Moreover, a final step
must be taken to verify the solution. This can be accomplished by checking if the slack variables are sufficiently small, i.e., \(b + b_{1} + b_{2} < b_{\text{min}}\) and \(b_{\text{min}}\) is set to ${10}^{-5}$ in the experiments.

\footnotetext{\({W}^{\star}(i)\) represents the \(i\)-th vector in the ordered set \({W}^{\star}\).}

\subsection{Complexity Reduction for Real-implementation} \label{sec:Time-domainComputationComplexityReduction}
Since applying range compression along fast time samples\cite{crockett2013introduction} (denoted as ``FFT+BPA'') can reduce the processing time of conventional BPA, we borrow its idea to further reduce the complexity of the proposed SAS approach (denoted as ``FFT+SAS''). Ignoring the noise term in \eqref{y_IF_n_m} and applying range-FFT to \(\mathbf{y}_{\text{IF},n}\), we have
\begin{eqnarray}
	Y_{\text{1D},n}(l) \!\!\!\!\!\!&=&\!\!\!\!\!\! \sum_{m = 0}^{M - 1}{y_{\text{IF},n}(m)e^{- j2\pi\frac{l}{L}m}}\nonumber \\
	\!\!\!\!\!\!&=&\!\!\!\!\!\! \alpha_{n}e^{j2\pi\left( \tau KT_{Start} + f_{c}\tau \right)}\!\!\!\sum_{m = 0}^{M - 1}e^{j2\pi\left( \tau Kt_{s} - \frac{l}{L} \right)m}\nonumber \\
	\!\!\!\!\!\!&=&\!\!\!\!\!\! \alpha_{n}e^{j4\pi\frac{KT_{Start} + f_{c}}{c}R_{n}}\!\!\!\sum_{m = 0}^{M - 1}e^{j2\pi\left( \frac{2R_{n}Kt_{s}}{c} - \frac{l}{L} \right)m},
\end{eqnarray}
where \(l = 0,1,\ldots,L - 1\) and \(L\) is the number of range
bins. Let \(k = \frac{2\pi\left( KT_{\text{Start}} + f_{c} \right)}{c}\). We
have
\begin{equation}
	Y_{\text{1D},n}(l) = \alpha_{n}e^{j2kR_{n}}\sum_{m = 0}^{M - 1}e^{j2\pi\left( \frac{2R_{n}Kt_{s}}{c} - \frac{l}{L} \right)m}, \label{Y_1D_n_l}
\end{equation}
and \(Y_{\text{1D},n}(l)\) reaches the maximum for
\begin{equation}
	l_{n}^{\star} = round\left( \frac{2R_{n}Kt_{s}L}{c} \right),\label{l_n_*}
\end{equation}
which is the range bin where the target is located. The data vector at the \(n\)-th phase center now
becomes
\begin{equation}
	\mathbf{y}_{\text{1D},n} = \left\lbrack Y_{\text{1D},n}(0),Y_{\text{1D},n}(1)\ldots,Y_{\text{1D},n}(L - 1) \right\rbrack^{T}.
\end{equation}
The data matrix from effective phase centers used for SAR is given by
\begin{equation}
	\mathbf{Y}_{\text{1D}} = \left\lbrack \mathbf{y}_{\text{1D},N_{\min}},\mathbf{y}_{\text{1D},N_{\min} + 1},\ldots,\mathbf{y}_{\text{1D},N_{\max}} \right\rbrack.
\end{equation}
To focus a point locating at \(\left( \phi_{t},R_{t} \right)\) in polar coordinates, we need to compute
\begin{equation}
	I( \phi_{t},R_{t} ) = \blue{\mathbf{w}^{H} \cdot \mathbf{Y}_{\text{1D}}( l_{N_{\min}}^{\star},l_{N_{\min} + 1}^{\star},\ldots,l_{N_{\max}}^{\star} )}, \label{I_1DFFT}
\end{equation}
where \(\mathbf{Y}_{\text{1D}}( l_{N_{\min}}^{\star},l_{N_{\min} + 1}^{\star},l_{N_{\min} + 2}^{\star},\ldots,l_{N_{\max}}^{\star} ) = \lbrack \mathbf{y}_{\text{1D},N_{\min}}( l_{N_{\min}}^{\star} ),\mathbf{y}_{\text{1D},N_{\min} + 1}( l_{N_{\min} + 1}^{\star} ),\ldots,\mathbf{y}_{\text{1D},N_{\max}}( l_{N_{\max}}^{\star} ) \rbrack^{T}\) and \(\mathbf{y}_{\text{1D},n}(l)\) represents the \(l\)-th entry of the vector \(\mathbf{y}_{\text{1D},n}\). 
\(\mathbf{w}\) in \eqref{I_1DFFT} can be the sparse weight vector of the corresponding range bin determined by Algorithm 2, or the weight vector of BPA with the \(j\)-th entry being \(\mathbf{w}_{\text{BP},j} = \alpha_{N_{\min} + j - 1}  \cdot e^{- j2kR_{N_{\min} + j - 1}}\).

\begin{remark}
	The steering vector may differ from the origin one in \eqref{OriginalArrayPattern} after applying range-FFT, since substituting \eqref{l_n_*} into \eqref{Y_1D_n_l} cannot fully cancel the phase summation term in some cases (e.g., when \(R_n\) is not multiple of the length of range bin). In our implementation, \(\mathbf{w}^{\star}\) is still obtained from the original steering vector. Thus, the array pattern could deviate from the desired one. However, doing so can lead to computation reduction.
\end{remark}
\begin{remark}
	The computation complexity of SAS is \(O( L_{x}\times L_{y}\times M\times N' )\), where \(N'\) is number of phased centers corresponding to non-zero weights. \(N'\) is typically less than a half of \(N\). As for FFT+SAS, the computation complexity is given by \(O( N'\times M\log_2 M + L_{x}\times L_{y}\times N' )\). When $L_x\times L_y \gg M$, the second term dominates. Thus, the overall reduction in complexity by combining FFT and SAS is substantial compared with that of the conventional BPA algorithm. Since the proposed algorithms conduct filtering pixel-by-pixel independently, they can be further accelerated by separating these pixels into multiple groups and processing them in a parallel manner.
\end{remark}

\section{Performance} \label{sec:SimulationStudy}
In this section, we conduct experimental study to evaluate the effectiveness of the proposed ROSAR imaging algorithms.

\subsection{Hardware and Software Implementation} \label{sec:HardwareandSoftwareImplementation}
\begin{figure}
	\centering
	\begin{subfigure}{0.367\linewidth}
		\includegraphics[width=\linewidth]{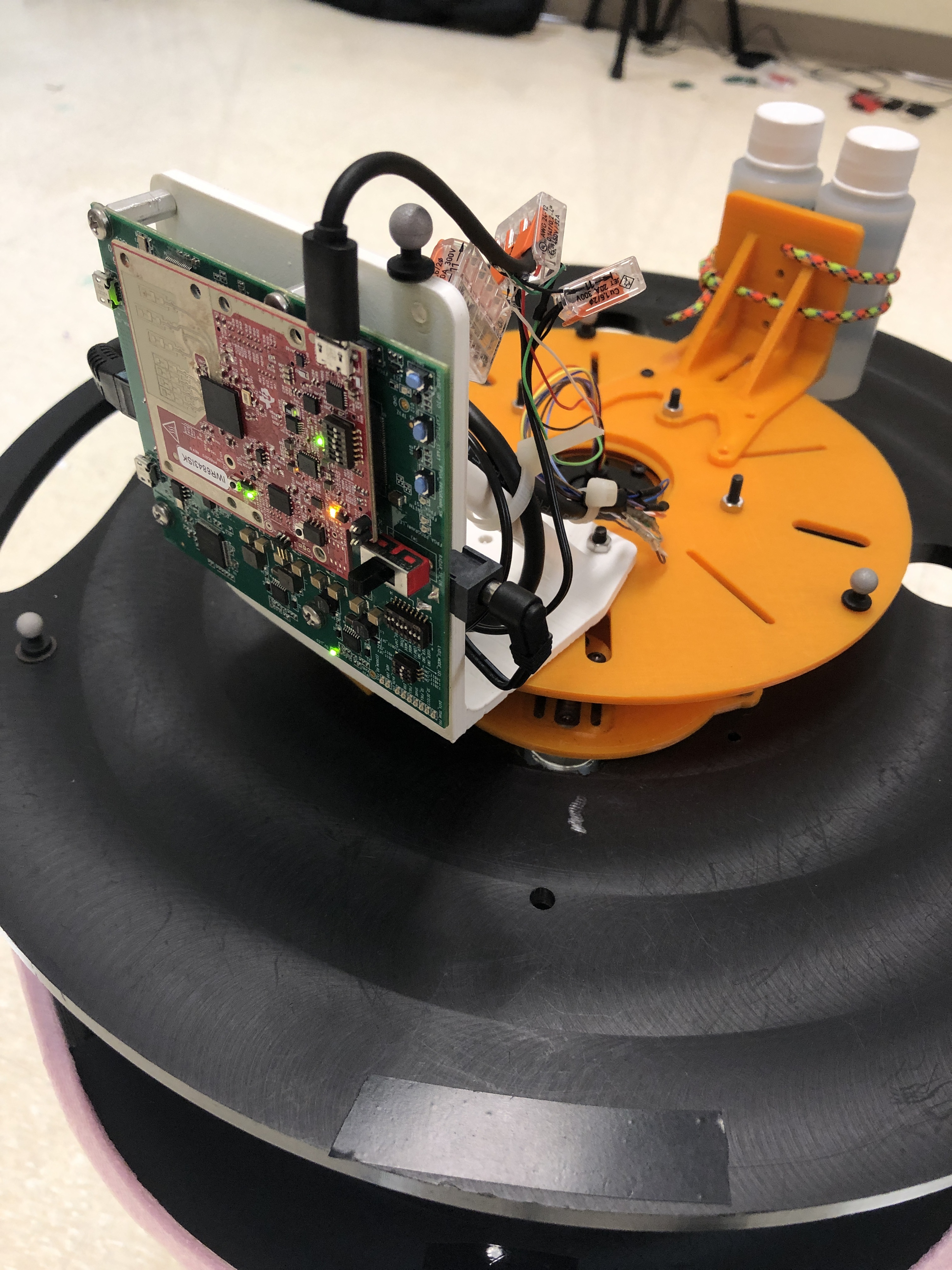}
	\end{subfigure}
	\begin{subfigure}{0.49\linewidth}
		\includegraphics[width=\linewidth]{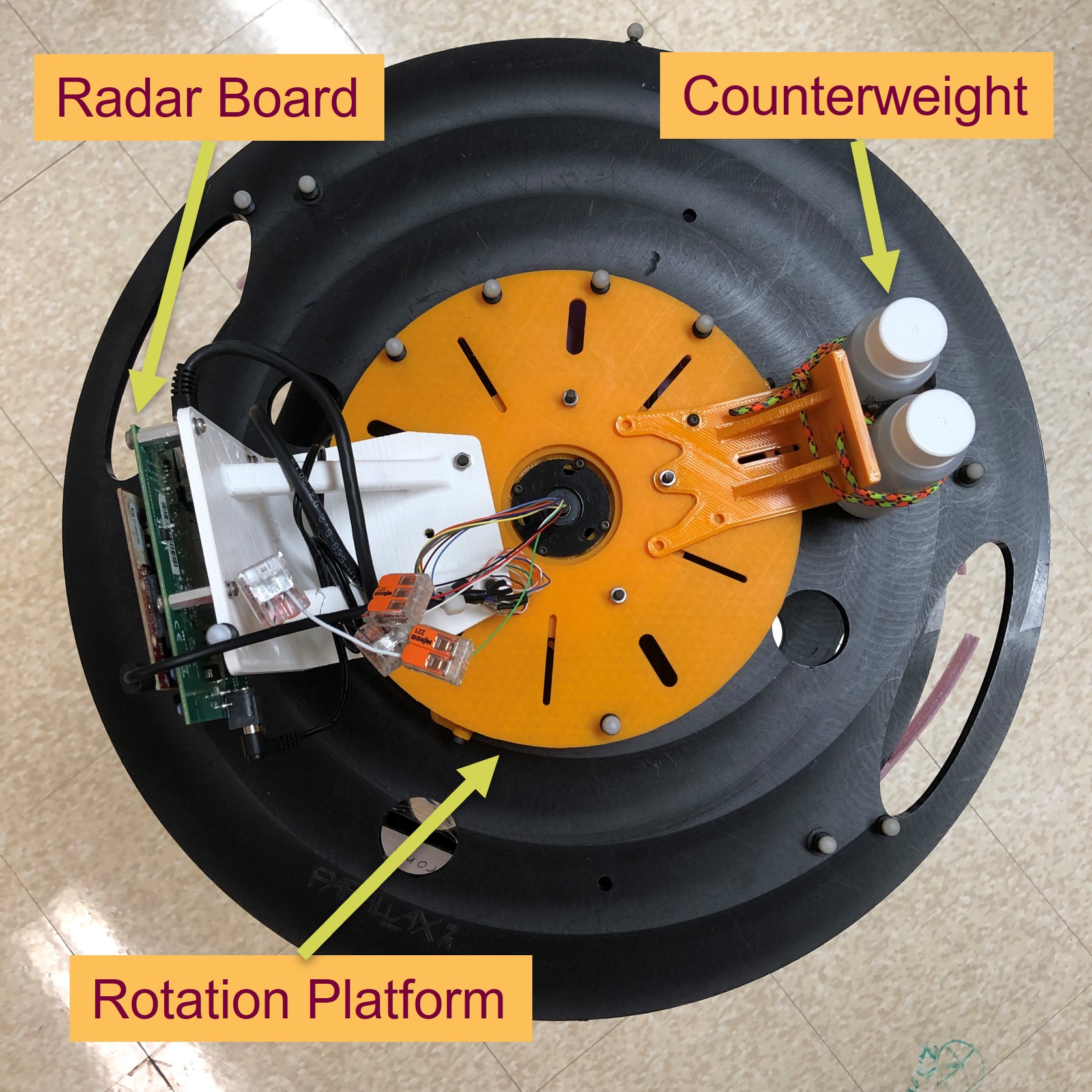}
	\end{subfigure}
	\caption{ROSAR system}
	\label{fig:ROSARSystem}
\end{figure}

We implement the proposed algorithms on a PC equipped with an Intel Core 8700 CPU and 16GB RAM. The PC has been installed MATLAB and Phased Array Toolbox. We also build a real ROSAR system, which hardware consists of a radar board, a rotation plate, motor and odometry sensors mounted on a rover platform as shown in Fig. \ref{fig:ROSARSystem}. The radar uses is Taxes Instruments IWR6843ISK, which generates millimeter wave (mmWave) signals. A 3D-printed plate holding the radar and counter weight is connected to a step motor and a wheel encoder.

The whole system is controlled by Robot Operating System (ROS), including the radar signal sending/receiving and rotation speed. Radar data is collected from the antenna board and angle reading is obtained from the encoder of the rotation platform. Both data is streamed to the PC and processed by MATLAB programs for SAR imaging. 

\begin{figure}
	\centering
	\includegraphics[width=0.66\linewidth]{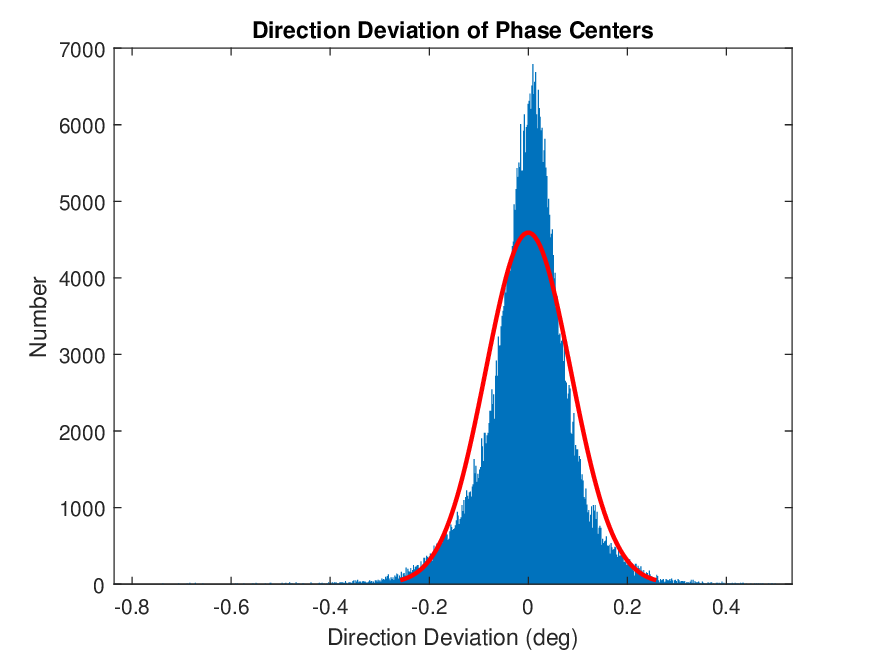}
	\caption{Direction deviation of phase centers}
	\label{fig:DirectionDeviation}
\end{figure}

To estimate phase center errors from wheel encoder, we mount several optical markers on the rotation plate. Markers can be tracked by OptiTrack, an optical motion capture system treated as ground truth (position accuracy: 0.2 mm). We rotate the platform multiple rounds and compare the angle difference between the data from tick's encoder and the OptiTrack system. Fig. \ref{fig:DirectionDeviation} shows the mean value is \(3.9 \times {10}^{-4}\) degree. If the data is fitted by a Gaussian distribution, the standard deviation is \(\sigma = 0.086\). Thus, we can assume the direction of the \(n\)-th phase center follows \({\widehat{\phi}}_{n}\sim\mathcal{N}\left( \frac{2\pi n}{N},0.086 \right)\).

\subsection{Parameter Settings}
\begin{figure}
	\centering
	\begin{subfigure}{0.49\linewidth}
		\includegraphics[width=\linewidth]{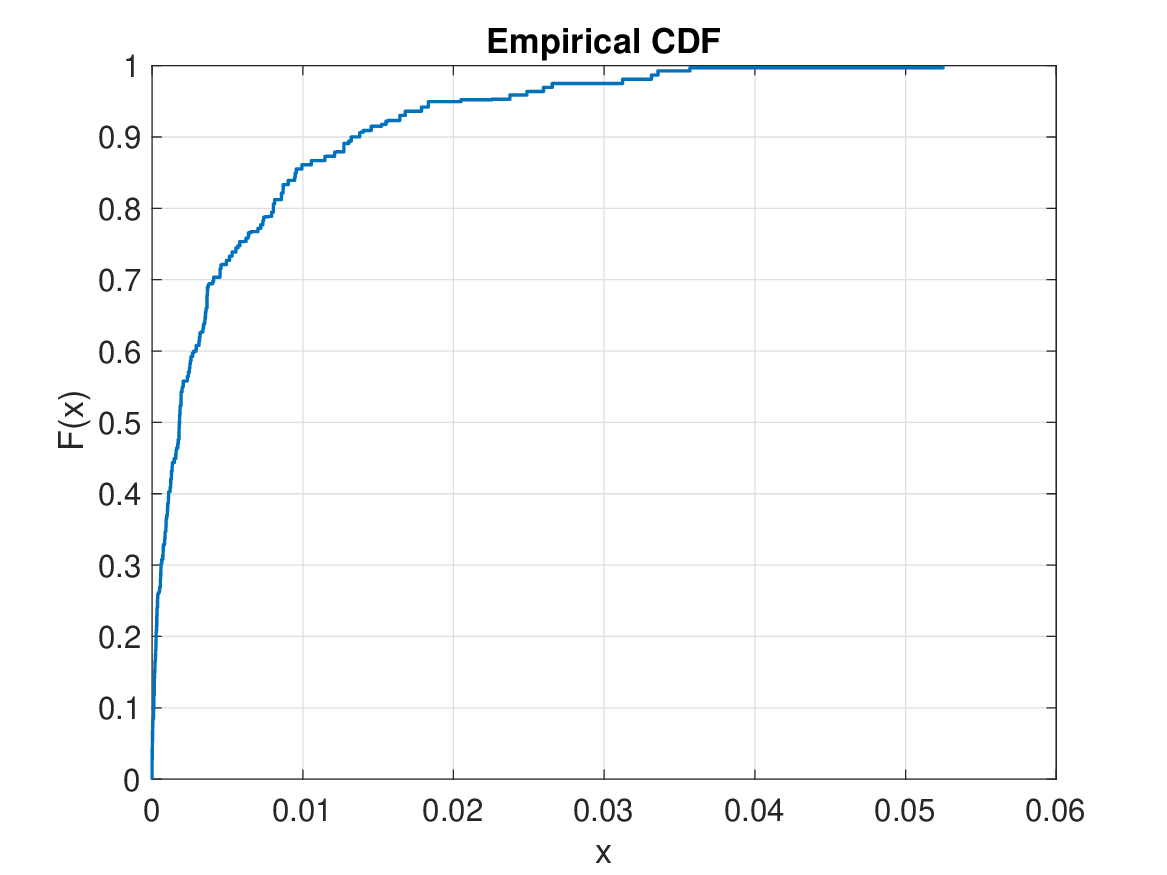}
		\caption{The CDF of $\widehat{\Delta}_{R}$ (\(R=2\))}
		\label{fig:DeltaCDF_R=2}
	\end{subfigure}
	\begin{subfigure}{0.49\linewidth}
		\includegraphics[width=\linewidth]{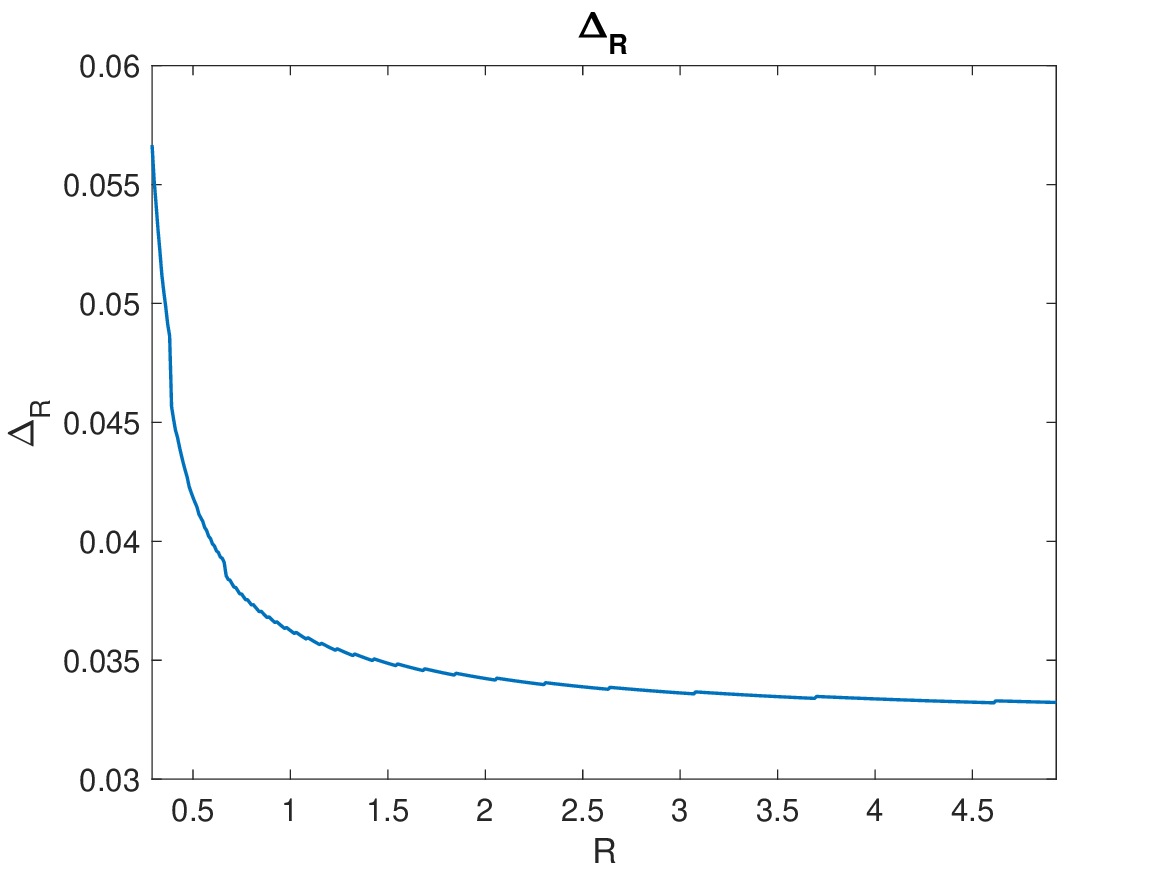}
		\caption{\({\Delta}_{R}\) for all range bins}
		\label{fig:Deltas}
	\end{subfigure}
	\caption{\({\Delta}_{R}\) calculation}
\end{figure}

The detailed parameters of the ROSAR system are summarized in Table \ref{tab:SimulationParameters}. From the settings, the range resolution is given by \(R_{\Delta} = \frac{c}{2B} = 0.0435m\), where \(B\) is
the bandwidth of the sampled chirp signal. The maximum unambiguous range
is \(R_{\max} = \frac{cF_{s}}{4K} \approx 4.8686m\). If we choose \(\Delta_{R}\) to represent 99\% of the
probability of the cumulative distribution function (CDF), Fig. \ref{fig:DeltaCDF_R=2} shows an example CDF for \(R=2\) and \(\Delta_{R} = 0.035\). Fig. \ref{fig:Deltas} shows \(\Delta_{R}\) as a function of range, from which we can see \({\Delta}_{R}\) decreases as the range becomes larger. It is because a small displacement from the desirable positions of phase centers has less impact when the radar is further away from the target. The detailed calculation steps have been given in Section \ref{ParameterSettings}.
\begin{table}
	\centering
	\caption {Parameters of the ROSAR System} \label{tab:SimulationParameters} 
	\begin{tabular}{|l|l|}
		\hline
		\multicolumn{2}{|c|}{Radar Settings}\\
		\hline
		\(r\) & 0.145 m \\
		\hline
		Rotation speed & 60 RPM \\
		\hline
		Rotation time & 1 s \\
		\hline
		\# Of TX & 1 \\
		\hline
		\# Of RX & 1 \\
		\hline
		Antenna Pattern & Cosine \\
		\hline
		Antenna FOV & {[}-90°, 90°{]} \\
		\hline
		Start rotating direction & 0° \\
		\hline
		Rotation direction & Counter-clockwise \\
		\hline
		TX power & 12 dBm \\
		\hline
		Antenna gain & 7 dBi \\
		\hline
		RX gain & 48 dB \\
		\hline
		\multicolumn{2}{|c|}{Chirp Signal Settings} \\
		\hline
		Start Frequency & 60 GHz \\
		\hline
		End Frequency & 64 GHz \\
		\hline
		Ramp start time & 0 us \\
		\hline
		Ramp end time & 58 us \\
		\hline
		\(T_{Start}\) & 7 us \\
		\hline
		Sampling end time & 57 us \\
		\hline
		\(F_{s}\) & 4.5 MHz \\
		\hline
		\(M\) & 225 \\
		\hline
		\(N\) & 800 \\
		\hline
		\(K\) & \(6.8 \times 10^{13}\) Hz/s\\
		\hline
		\(L\) & 225 \\
		\hline
	\end{tabular}
\end{table}

All the sparse weight vector \(\mathbf{w}\) for each range bin are computed
in advance using Algorithm \ref{algo:Pre-Compute} with parameters listed in Table \ref{tab:OptimizationProblemParameters}. We use CVX and Mosek solver\cite{cvx} to find the optimal values in each iteration of SCA.
\begin{table}
	\centering
	\caption {SAS Parameters} \label{tab:OptimizationProblemParameters} 
	\begin{tabular}{|l|l|}
		\hline
		\(ITER\) & 50 \\
		\hline
		\(Th\) & 0.001 \\
		\hline
		\(\phi_{\text{MW}}\) & 1\degree \\
		\hline
		Sampled angle interval & 0.5\degree \\
		\hline
		\(\lambda_{b}\) & 50 \\
		\hline
		\(\eta\) & 0.0005 (-33 dB) \\
		\hline
		\(U_{\min}\) & 5\\
		\hline
	\end{tabular}
\end{table}

\subsection{Baseline Algorithm and Metrics}
We implement four algorithms: ``BPA'', ``FFT+BPA'', ``SAS''\blue{, ``FFT+SAS'', ``RBPA'' (BPA with randomly selected phase centers) and ``FFT+RBPA''} for different comparison purposes. The following metrics are used in quantitative evaluations:
\begin{itemize}
	\item \textit{Half main lobe width} \(\phi_{\text{MW}}\): The half main lobe width is defined as the angle interval between the peak and the closest local minima on either side of the main lobe.
	\item \textit{Peak-to-integral sidelobe ratio (PISR)}: The PISR
	\(\mathcal{R}\) for a specific range bin \(R\) is calculated as
\end{itemize}
\begin{equation}
	\mathcal{R} = \frac{\left| I\left( \phi_{m},R \right) \right|^{2}}{\sum_{s = 1}^{S}\left| I\left( \phi_{s},R \right) \right|^{2}}.
\end{equation}
\begin{itemize}
	\item \textit{SAR computation cost}: The elapsed time of generating a SAR image.
	\item \textit{Image entropy} \cite{xu2011weighted}: Let
	\(E = \sum_{\phi}^{}{\sum_{R}^{}\left| I(\phi,R) \right|^{2}}\) be the
	total energy of the image, and \(d_{(\phi,R)} = \frac{\left| I(\phi,R) \right|^{2}}{\ E = \sum_{\phi}^{}{\sum_{R}^{}\left| I(\phi,R) \right|^{2}}}\) be the energy density of a pixel. The image entropy is defined as
\end{itemize}
\begin{equation}
	E_{I} = - \sum_{\phi,R}^{}{d_{(\phi,R)}\ln d_{(\phi,R)}}.
\end{equation}

The targets are well focused on the SAR image if \(\mathcal{R}\) is large and \(E_{I}\) is small.

\subsection{Numerical Results and Analysis}\label{ResultComparisonandAnalysis}
\subsubsection{Solution to the Optimization Problem}
\begin{figure}
	\begin{subfigure}{0.49\linewidth}
		\includegraphics[width=\linewidth]{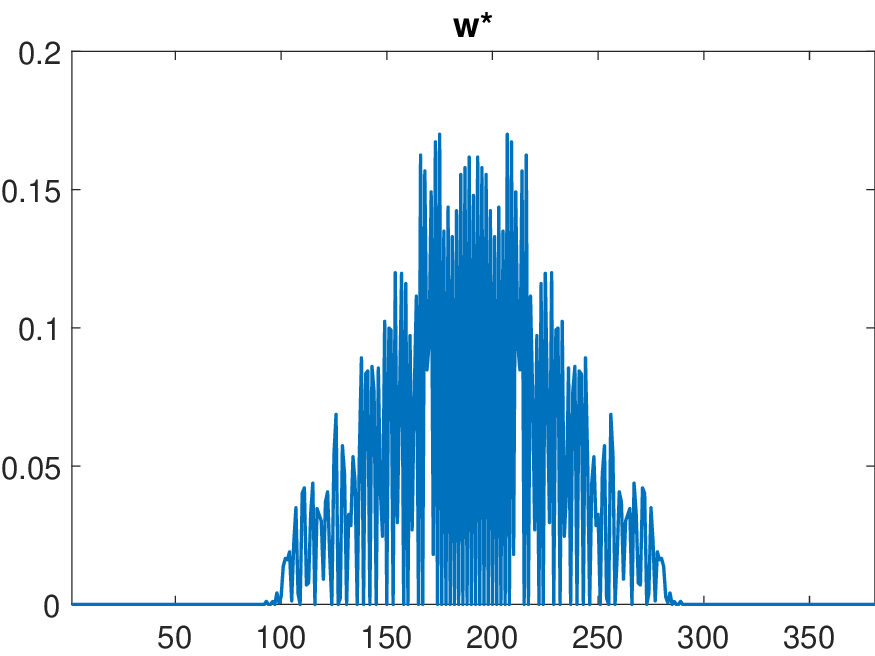}
		\caption{Non-robust design}
	\end{subfigure}
	\begin{subfigure}{0.49\linewidth}
		\includegraphics[width=\linewidth]{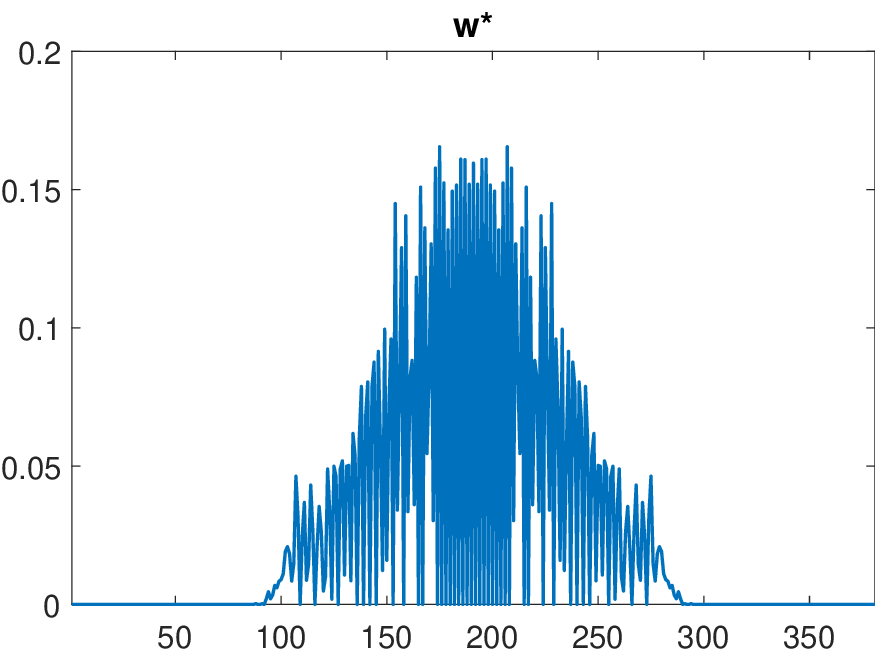}
		\caption{Robust design}
	\end{subfigure}
	\caption{The magnitude of each element in \(\mathbf{w}^{*}\) under different scenarios}
	\label{fig:w*}
\end{figure}

We first give the numerical result through imaging a point target locating at \(\left( \frac{\pi}{2},2m \right)\). In this case, the number of effective phase centers, i.e., the length of synthesized aperture, is calculated to be 381 by \eqref{n}. Fig. \ref{fig:w*} shows the magnitude of each element in \(\mathbf{w}^{\star}\) out and  robust design, respectively. In both cases, we can see \(\left\| \mathbf{w} \right\|_{0}\) is less than a half of the total number of phase centers from Table \ref{tab:Nw&U'}. However,  the robust design, the sparsity is slightly reduced.
\begin{table}
	\centering
	\caption {Solution to the Optimization Problem} \label{tab:Nw&U'} 
	\begin{tabular}{|l|c|c|}
		\hline
		& $\left\| \mathbf{w} \right\|_{0}$ & $U'$\\
		\hline
		Non-Robust (\(\Delta_{R} = 0\)) & 141 & 8.21 \\
		\hline
		Robust (\(\Delta_{R} = 0.035\)) & 157 & 8.38 \\
		\hline
	\end{tabular}
\end{table}

\subsubsection{Array Pattern}
\begin{figure}
	\centering
	\begin{subfigure}{0.49\linewidth}
		\includegraphics[width=\linewidth]{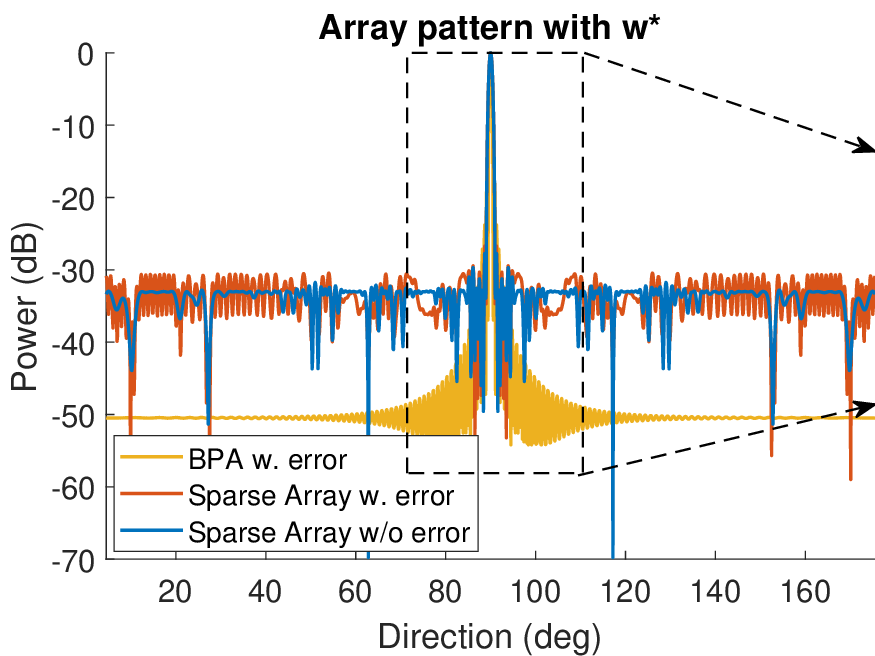}
		\caption{Non-robust design}
		\label{fig:ArrayPattern_R=2_NonRobust}
	\end{subfigure}
	\begin{subfigure}{0.49\linewidth}
		\includegraphics[width=\linewidth]{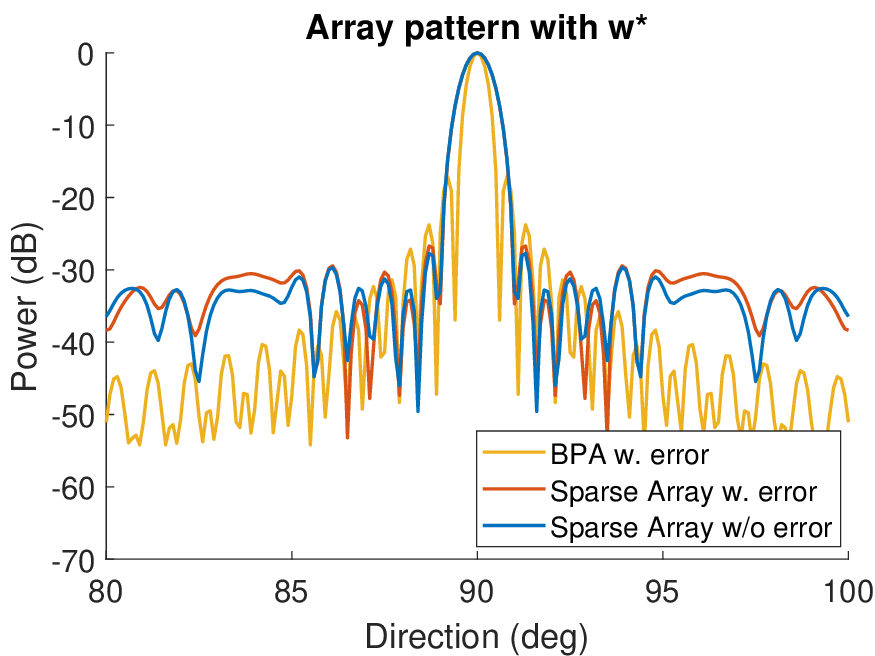}
		\caption{Main lobe region of (a)}
		\label{fig:ArrayPattern80100_R=2_NonRobust}
	\end{subfigure}
	\begin{subfigure}{0.49\linewidth}
		\includegraphics[width=\linewidth]{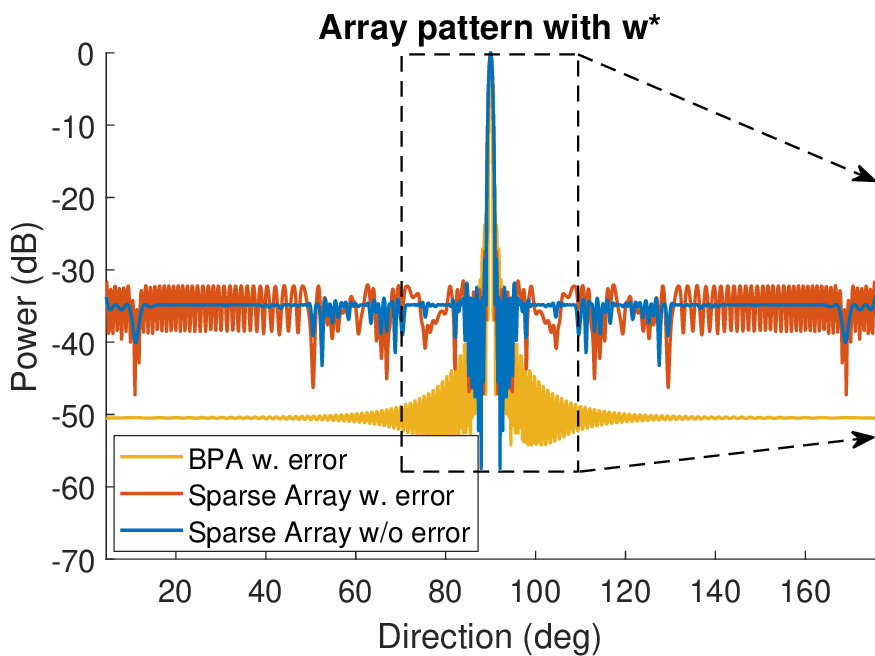}
		\caption{Robust design}
		\label{fig:ArrayPattern_R=2_Robust}
	\end{subfigure}
	\begin{subfigure}{0.49\linewidth}
		\includegraphics[width=\linewidth]{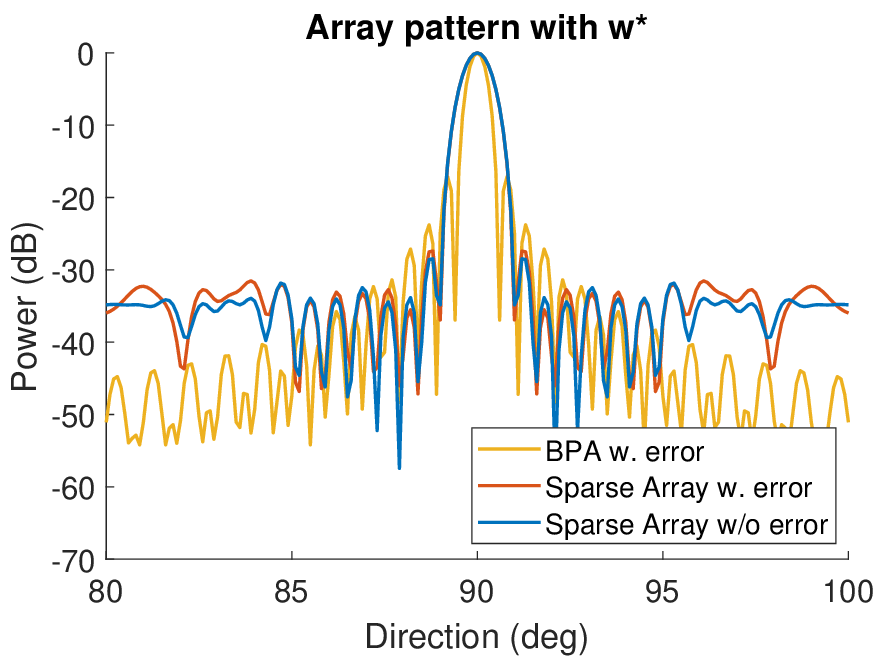}
		\caption{Main lobe region of (c)}
		\label{fig:ArrayPattern80100_R=2_Robust}
	\end{subfigure}
	\caption{The Array Pattern in Different Scenarios}
	\label{fig:ArrayPattern_R=2}
\end{figure}

Fig. \ref{fig:ArrayPattern_R=2} shows the array patterns of the SAS out and  considering
robust design for \(R=2\). Fig. \ref{fig:ArrayPattern80100_R=2_NonRobust} and Fig.  \ref{fig:ArrayPattern80100_R=2_Robust} are the main lobe area of Fig.~\ref{fig:ArrayPattern_R=2_NonRobust} and \ref{fig:ArrayPattern_R=2_Robust}. The array pattern is calculated by
\(F(\phi;2) = {\mathbf{w}^{\star H}}\mathbf{a}(\phi;2)\) for
\(\phi = \left\lbrack \frac{\pi}{2} - \phi_{v},\frac{\pi}{2} + \phi_{v} \right\rbrack\).
The blue and red lines show the array pattern out and
adding phase errors \(\mathbf{e}\), respectively. The yellow line is the array pattern of BPA  phase errors. As we can see,
the main lobe is in \(\lbrack - 1{^\circ},1{^\circ}\rbrack\),
which meets the design parameters. As shown in Fig. \ref{fig:ArrayPattern_R=2_NonRobust}, the power of the sidelobes when there is no error \(\mathbf{e}\) in the steering vector is mostly -33dB lower than that of the main lobe peak. The few exceptions fall in between angle grid points of interval 0.5° and thus their power levels are not enforced by the constraints.
Although denser grid points (and consequently more constraints) can reduce the chance of requirement violation, the computation cost of SCA grows drastically. When there are errors in the steering vector, most of the sidelobes do
not meet the -33dB criteria in non-robust design. In contrast,  the robust
design (the red line in Fig. \ref{fig:ArrayPattern_R=2_Robust}), this is no longer the case. Furthermore, the average power of the sidelobes is lower than -33dB  robust design even in absence of steering vector errors. This is due to the worst case assumption of robustness design as evident in \eqref{C1squareroot} and \eqref{C2squareroot}. The PISRs are given in Table \ref{tab:Peak-to-IntegralSidelobeRatio}. High values are better. The conventional BPA gives the best PISR due to its low sidelobes but needs much more computation time.
\begin{table}
	\centering
	\caption {Peak-to-Integral Sidelobe Ratio} \label{tab:Peak-to-IntegralSidelobeRatio} 
	\begin{tabular}{|c|c|c|}
		\hline
		Algorithms \& Settings & w/o error & w. error\\
		\hline
		SAS w. Robust & 0.1256 & 0.1253 \\
		\hline
		SAS w/o robust & 0.1308 & 0.1303 \\
		\hline
		BPA w. error & N/A & 0.2339 \\
		\hline
	\end{tabular}
\end{table}

\subsubsection{Results for All Range Bins}
\begin{figure}
	\begin{subfigure}{0.5\linewidth}
		\includegraphics[width=\linewidth]{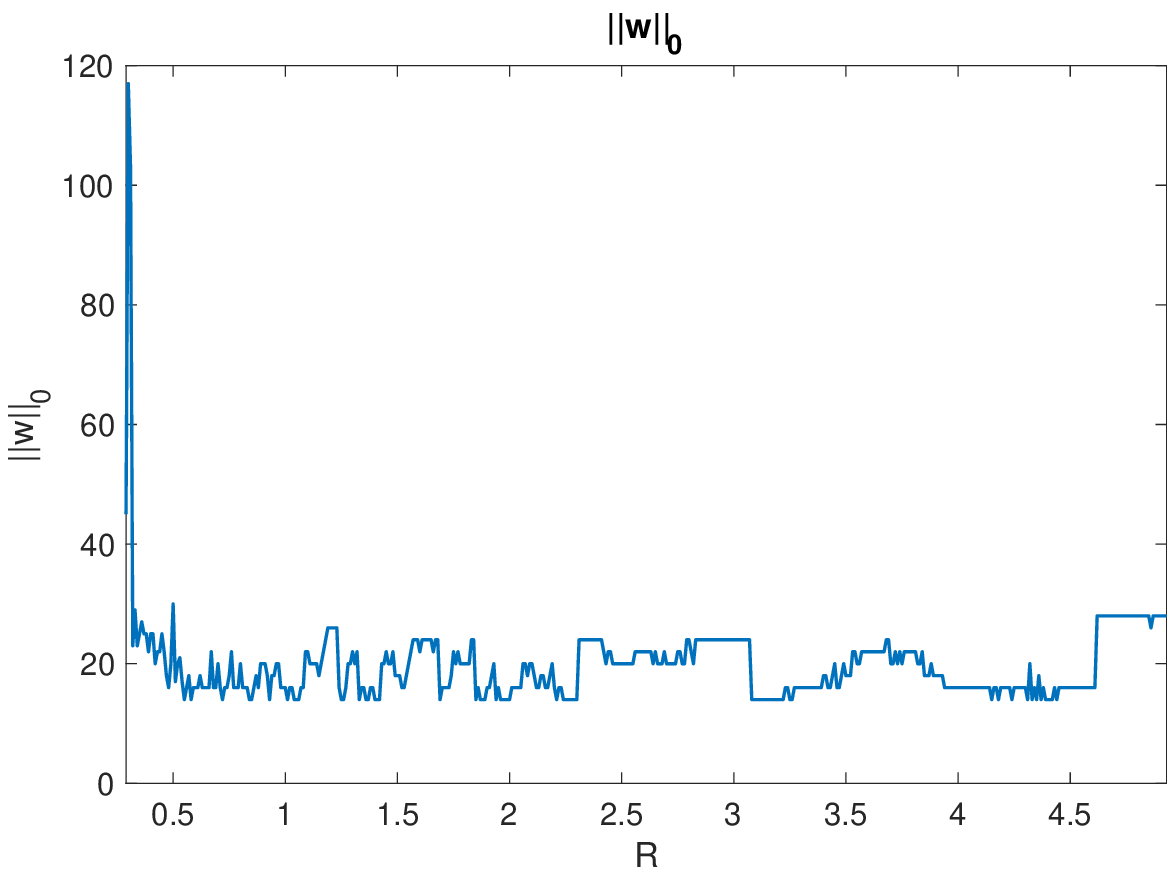}
		\caption{\(\left\| \mathbf{w} \right\|_{0}\)}
		\label{fig:Nws_Uge0}
	\end{subfigure}
	\begin{subfigure}{0.48\linewidth}
		\includegraphics[width=\linewidth]{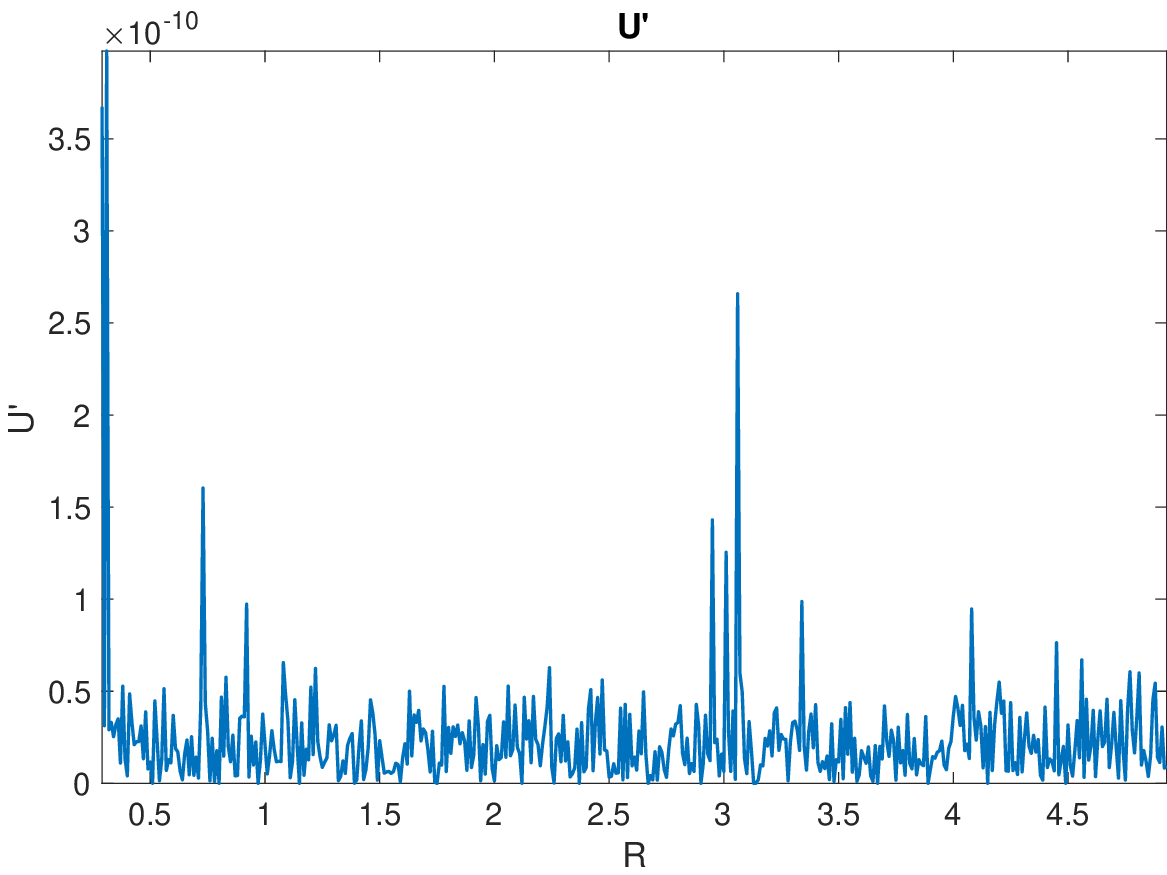}
		\caption{\(U'\)}
		\label{fig:U's_Uge0}
	\end{subfigure}
	\caption{The solutions of the SCA  \(U_{\min} = 0\) for different range bins}
	\label{fig:Results_R=all_Uge0}
\end{figure}

\begin{figure}
	\begin{subfigure}{0.5\linewidth}
		\includegraphics[width=\linewidth]{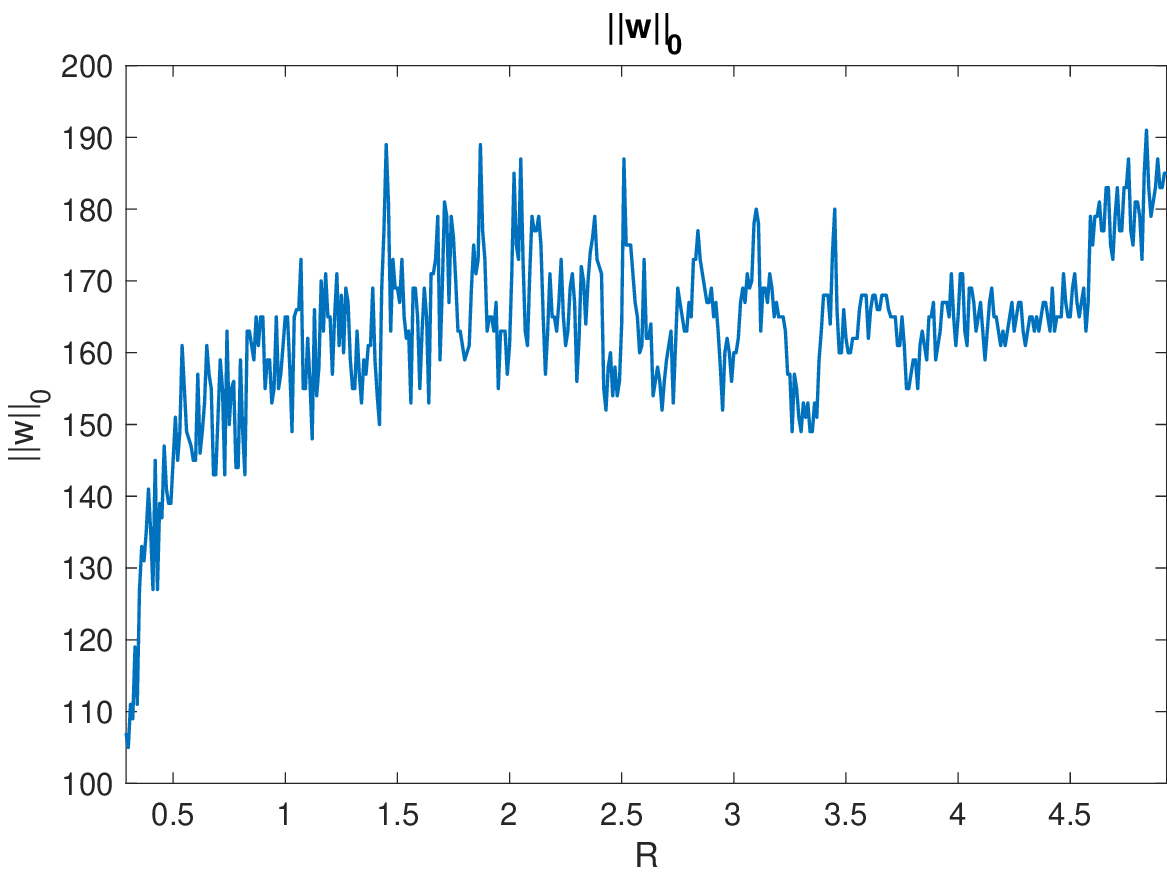}
		\caption{\(\left\| \mathbf{w} \right\|_{0}\)}
		\label{fig:Nws_Uge5}
	\end{subfigure}
	\begin{subfigure}{0.48\linewidth}
		\includegraphics[width=\linewidth]{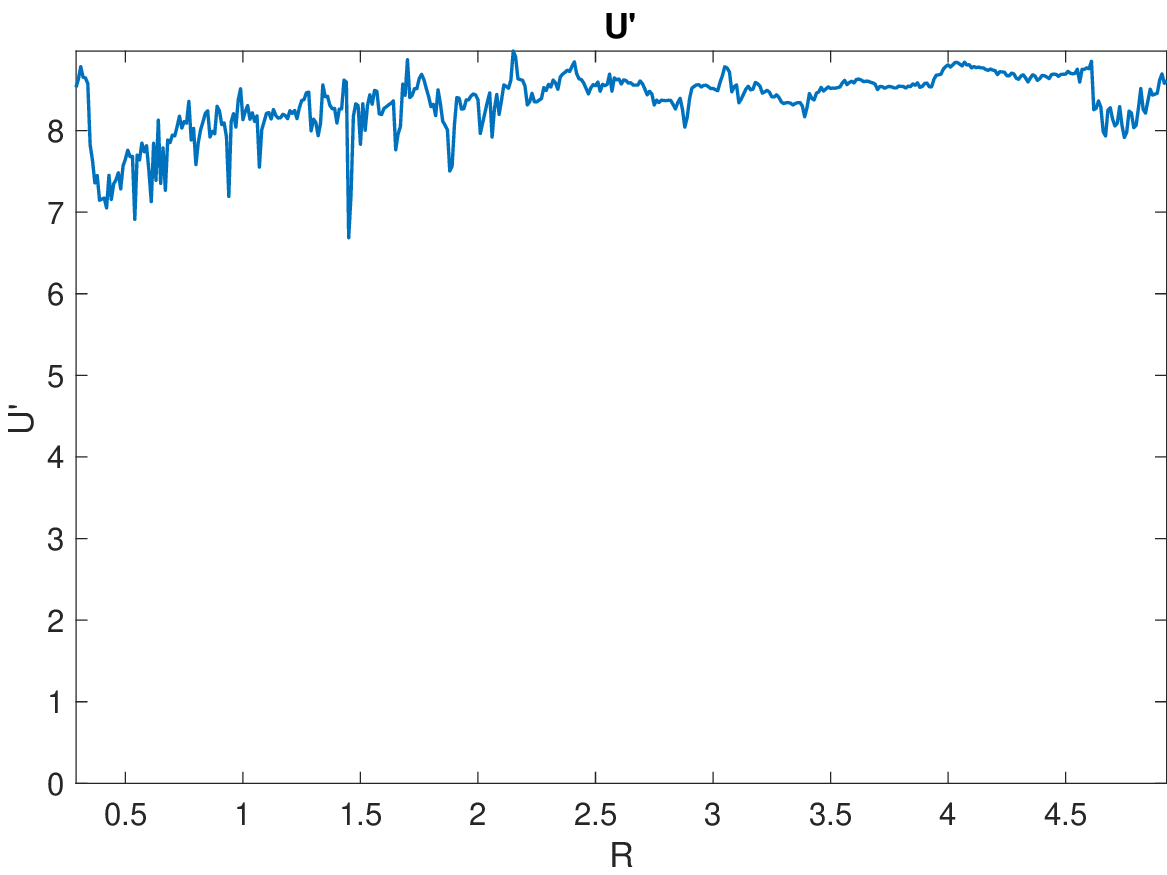}
		\caption{\(U'\)}
		\label{fig:U's_Uge5}
	\end{subfigure}
	\caption{The solutions of the SCA  \(U_{\min} = 5\) for different range bins}
	\label{fig:Results_R=all_Uge5}
\end{figure}

Fig. \ref{fig:Results_R=all_Uge0} shows the values of \(\left\| \mathbf{w} \right\|_{0}\) and \(U'\) for all range bins when $U_{\text{min}} = 0$. Clearly, the sparsity of the array holds in all range bins. We observe that \blue{all $U'$s are small. In this case}, although the array is very sparse, the SNR is low (recall that $U'$ is the magnitude of main-lobe peak and the noise power is constant from C3). Setting $U_{\text{min}} = 5$ can bound the SNR at the cost of reduced sparsity as shown in Fig. \ref{fig:Results_R=all_Uge5}.

\subsection{SAR Imaging Simulation for a Point Target}
\begin{figure}
	\centering
	\begin{subfigure}{0.49\linewidth}
		\includegraphics[width=\linewidth]{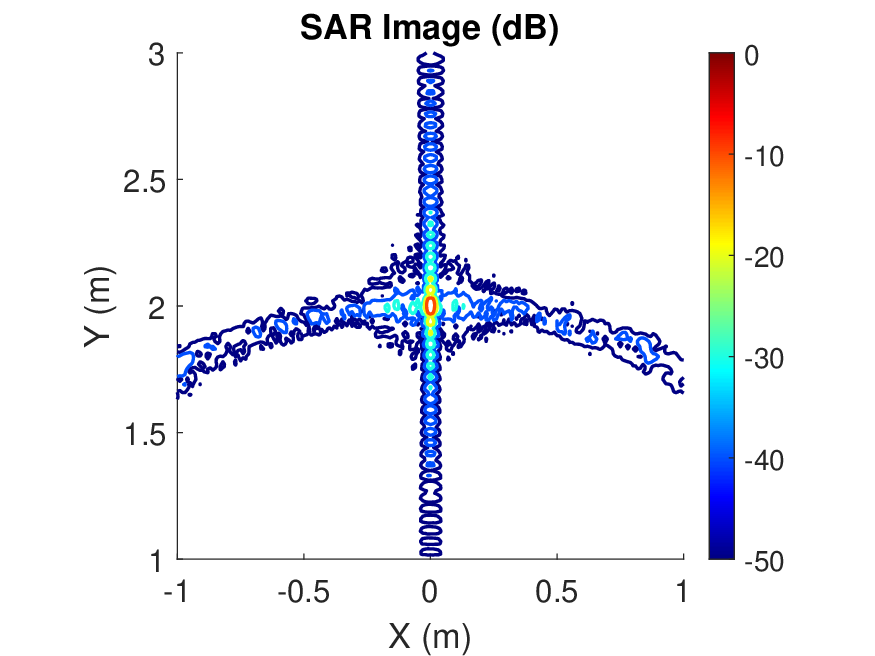}
		\caption{BPA}
		\label{fig:SARContourMapdB_BPA}
	\end{subfigure}
	\begin{subfigure}{0.49\linewidth}
		\includegraphics[width=\linewidth]{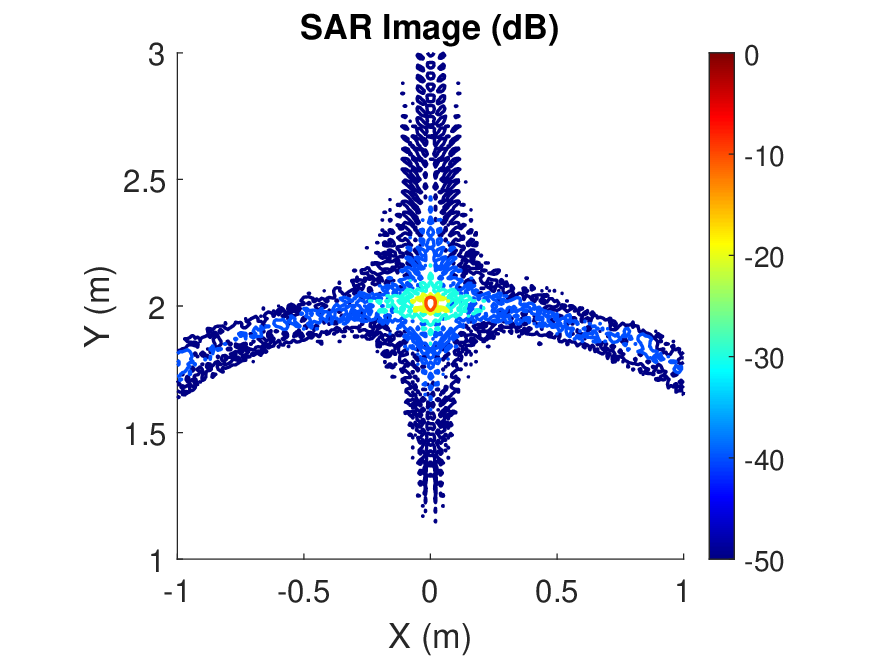}
		\caption{FFT + BPA}
		\label{fig:SARContourMapdB_FFT+BPA}
	\end{subfigure}\\
	\begin{subfigure}{0.49\linewidth}
		\includegraphics[width=\linewidth]{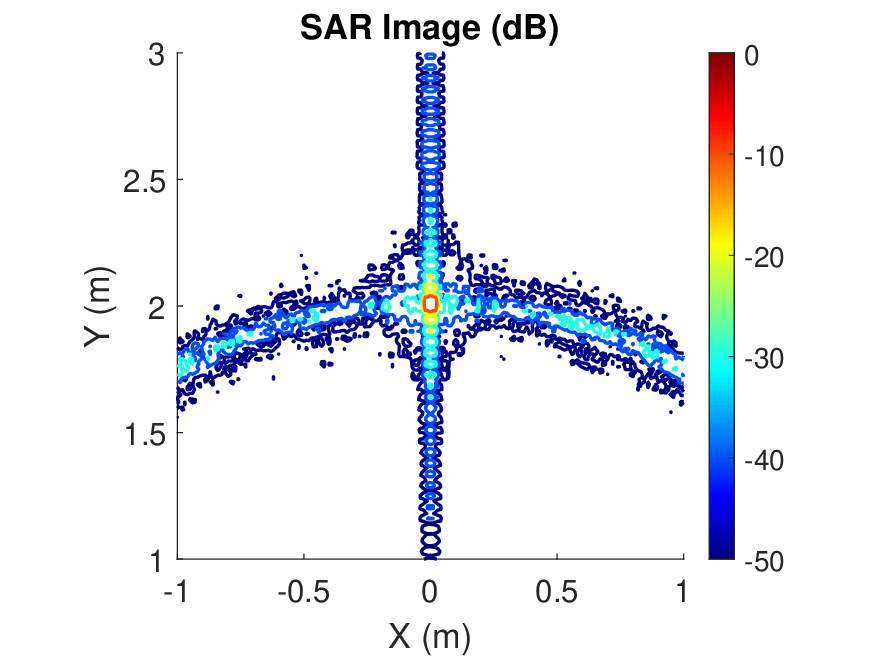}
		\caption{\blue{SAS}}
		\label{fig:SARContourMapdB_wle1_Uge5_loop=51}
	\end{subfigure}
	\begin{subfigure}{0.49\linewidth}
		\includegraphics[width=\linewidth]{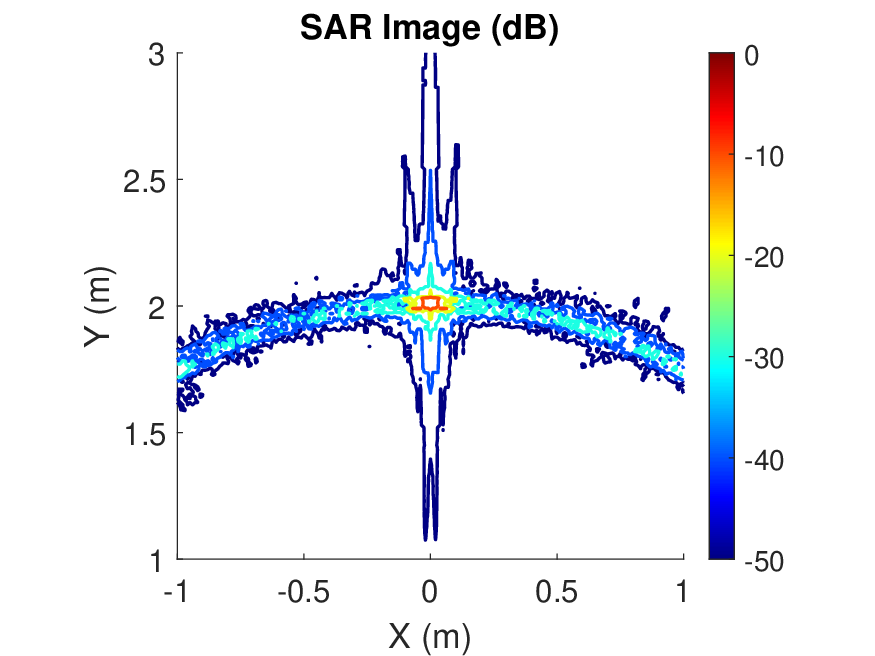}
		\caption{\blue{FFT + SAS}}
		\label{fig:SARContourMapdB_wle1_Uge5_loop=51_FFT}
	\end{subfigure}\\
	\begin{subfigure}{0.49\linewidth}
		\includegraphics[width=\linewidth]{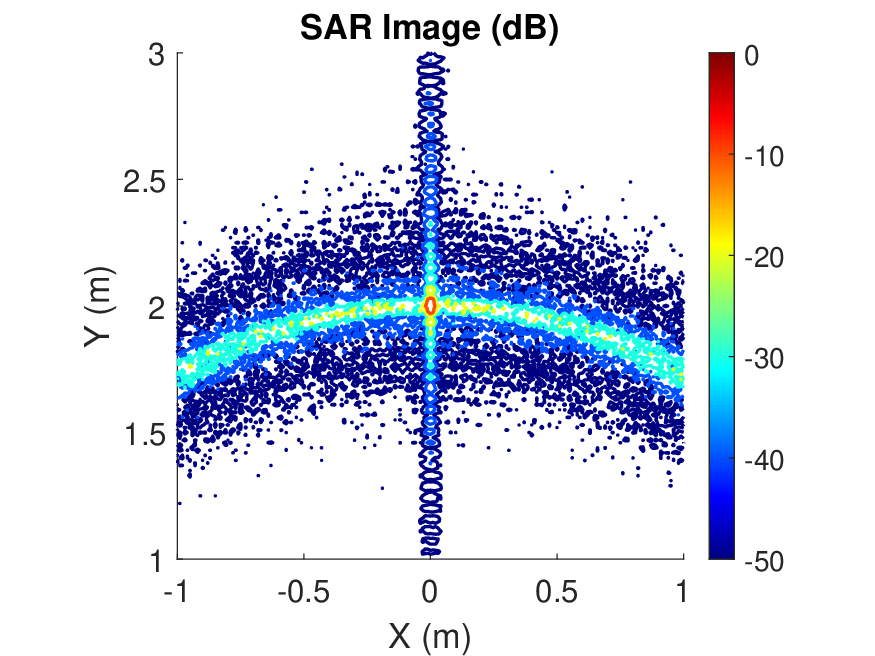}
		\caption{\blue{RBPA}}
		\label{fig:SARContourMapdB_BPA_Random}
	\end{subfigure}
	\begin{subfigure}{0.49\linewidth}
		\includegraphics[width=\linewidth]{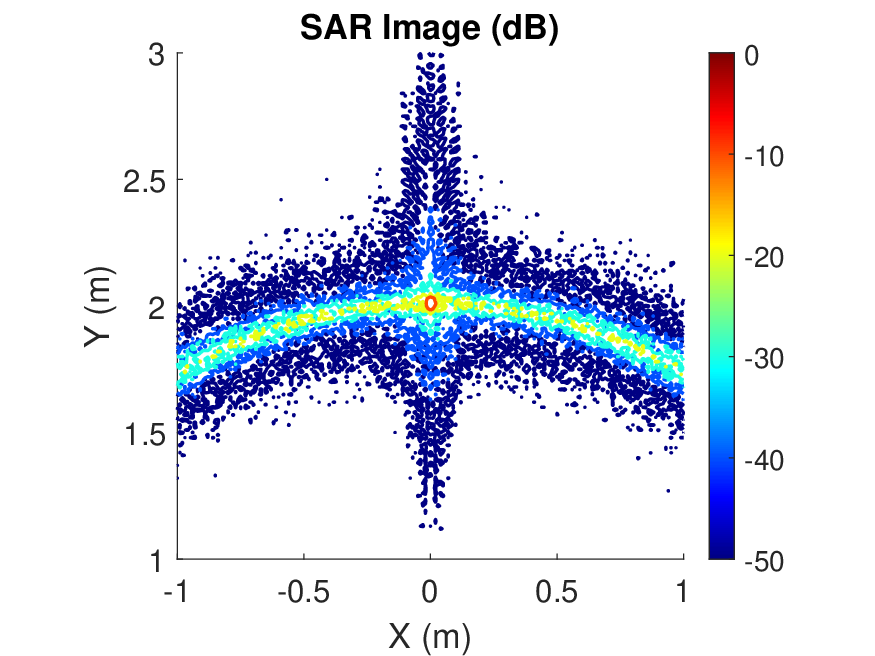}
		\caption{\blue{FFT + RBPA}}
		\label{fig:SARContourMapdB_FFT+BPA_Random}
	\end{subfigure}
	\caption{The SAR Image (Contour) of a point at \(\left( \frac{\pi}{2},2m \right)\)}
	\label{fig:SARMap_Sim_020}
\end{figure}

By using MATLAB Phased Array Toolbox, we simulated a point target locating at \(\left( \frac{\pi}{2},2m \right)\), a rotating radar and the sending/receiving signals of radar antennas. Fig. 
\ref{fig:SARContourMapdB_BPA}, \ref{fig:SARContourMapdB_wle1_Uge5_loop=51} and \ref{fig:SARContourMapdB_BPA_Random} show the imaging results of the target area by conventional BPA, SAS and RBPA, while Fig. \ref{fig:SARContourMapdB_FFT+BPA}, \ref{fig:SARContourMapdB_wle1_Uge5_loop=51_FFT} and \ref{fig:SARContourMapdB_FFT+BPA_Random} show the SAR images by employing FFT acceleration. The SAR image quality and computation cost is summarized in Table \ref{tab:ImageAPointTargetSimulation}. Although the entropy of the SAR image generated by SAS is slightly worse than that by BPA, the computational time is significantly reduced. \blue{Furthermore, although BPA with randomly selected phase centers takes less time as well, the resulting image quality is much worse than others.}
\begin{table}
	\centering
	\caption {SAR image quality and time cost} \label{tab:ImageAPointTargetSimulation} 
	\begin{tabular}{|c|c|c|}
		\hline
		Algorithm & $E_{I}$ & Time Cost (s)\\
		\hline
		BPA & \textbf{3.7489} & 43.66 \\
		\hline
		FFT + BPA & 3.9812 & 29.43 \\
		\hline
		SAS & \blue{4.4073} & \blue{18.03} \\
		\hline
		FFT + SAS & \blue{4.4703} & \blue{\textbf{3.61}} \\
		\hline
		\blue{RBPA} & \blue{5.7850} & \blue{18.03} \\
		\hline
		\blue{FFT + RBPA} & \blue{5.8728} & \blue{5.86} \\
		\hline
	\end{tabular}
\end{table}

\subsection{Testbed Evaluation} \label{sec:Testbed Evaluation}
Although simulations can provide insights on the impacts of configuration parameters and the performance of the proposed approach in simulated environments, existing packages in MATLAB cannot model the reflection, diffusion and deflection properties of mmWave signals in indoor environments well. In this section, the performance and efficiency of the proposed approach is validated through two real-world experiments. The size of SAR area is set to be 9.8756m\(\times\)9.8756m with grid size being 0.04m\(\times\)0.04m.

\subsubsection{Scenario 1: Corner Reflector}
\begin{figure}
	\centering
	\begin{subfigure}{0.46\linewidth}
		\includegraphics[width=\linewidth]{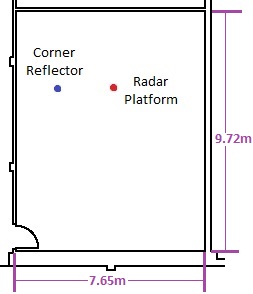}
		\caption{Geometry Relation}
	\end{subfigure}
	\begin{subfigure}{0.4\linewidth}
		\includegraphics[width=\linewidth]{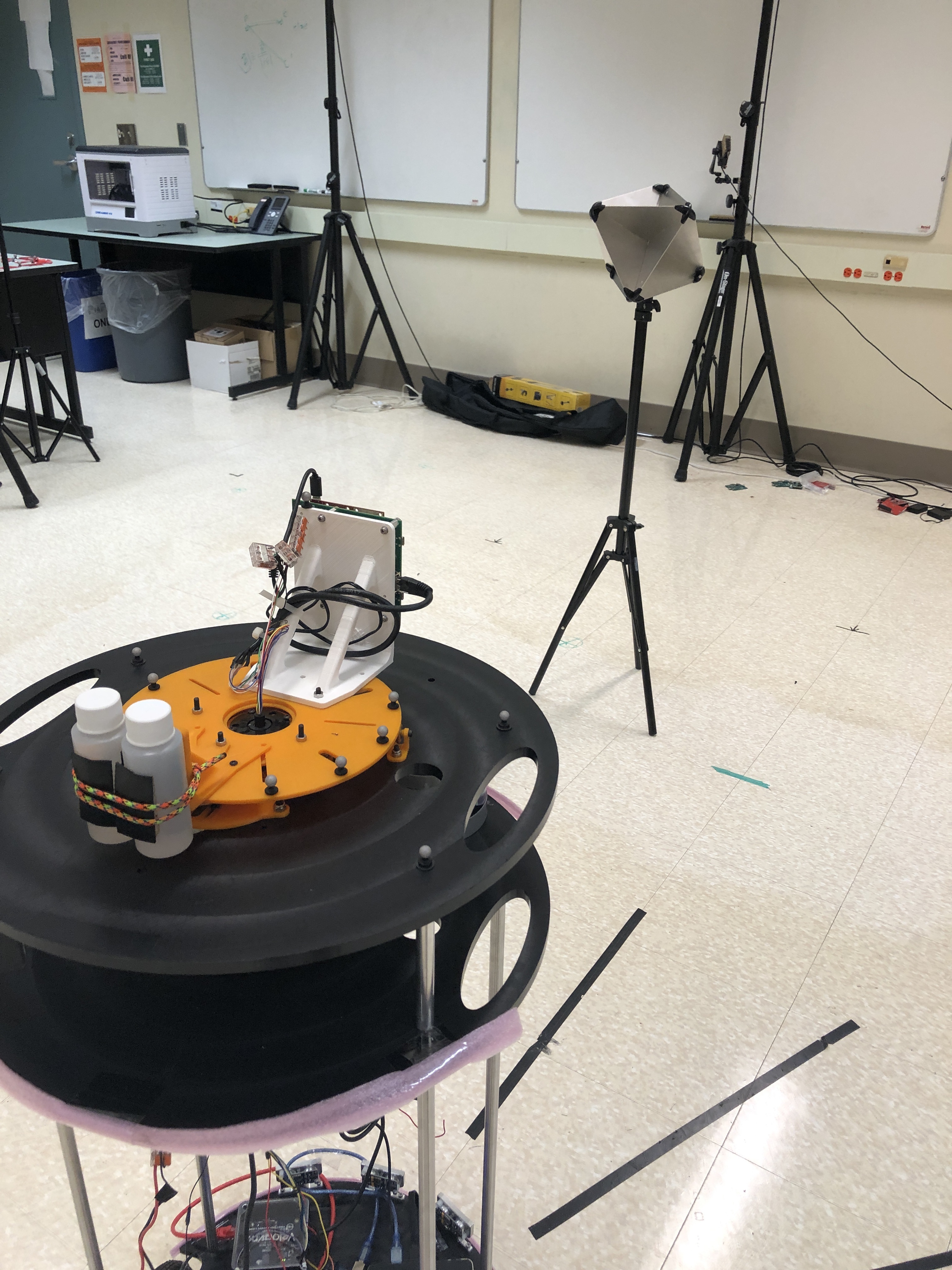}
		\caption{Experimental Scenario}
	\end{subfigure}
	\caption{Experiment setup for imaging a corner reflector}
	\label{fig:CornerReflector}
\end{figure}

\begin{figure}
	\begin{subfigure}{0.49\linewidth}
		\includegraphics[width=\linewidth]{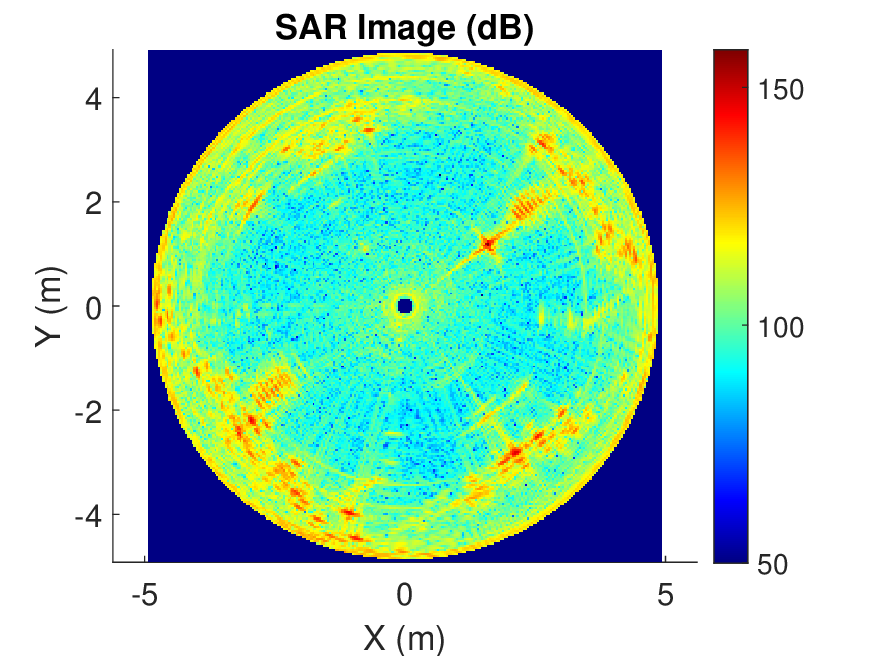}
		\caption{Back-projection\\algorithm}
	\end{subfigure}
	\begin{subfigure}{0.49\linewidth}
		\includegraphics[width=\linewidth]{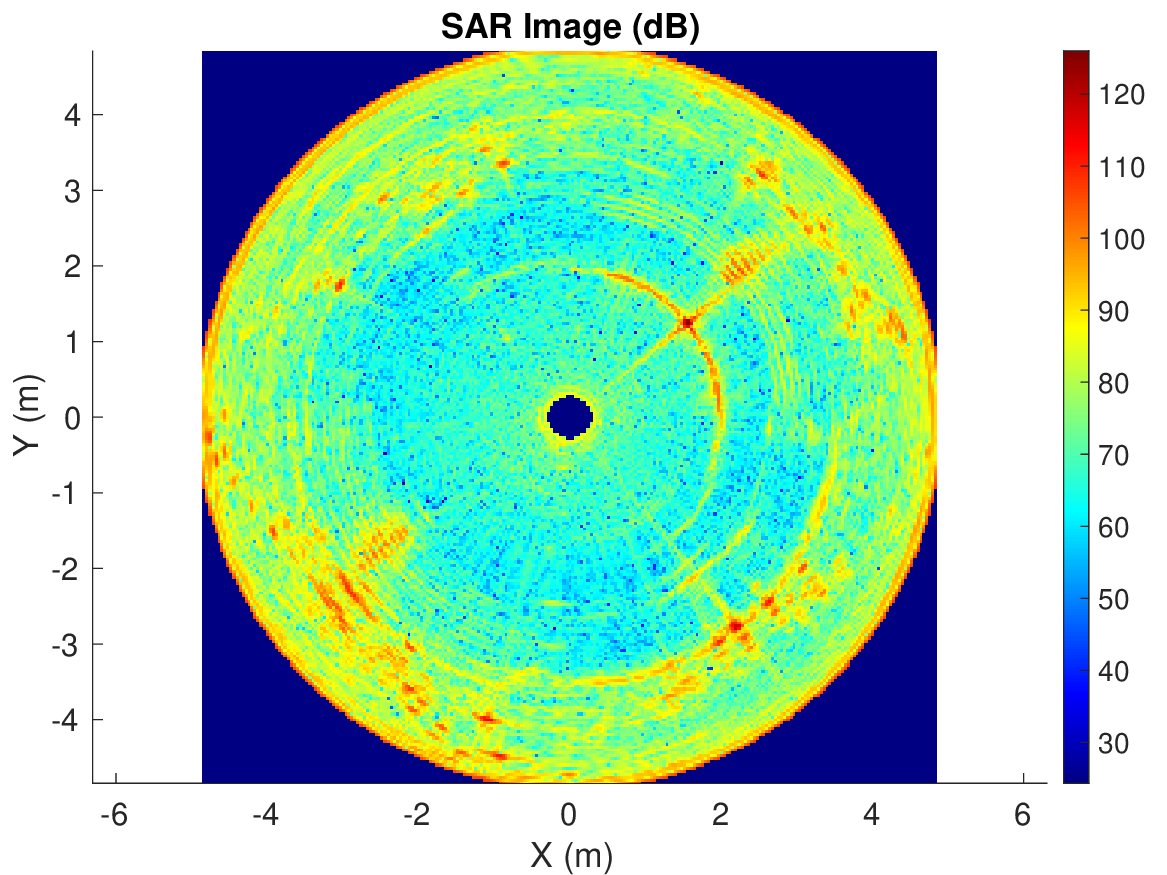}
		\caption{\blue{SAS, \(\phi_{\text{MW}}\!\!=\!\!1\degree\),\(\eta\!\!=\!\!-30\)dB,w/o Robust}}
	\end{subfigure}
	\begin{subfigure}{0.49\linewidth}
		\includegraphics[width=\linewidth]{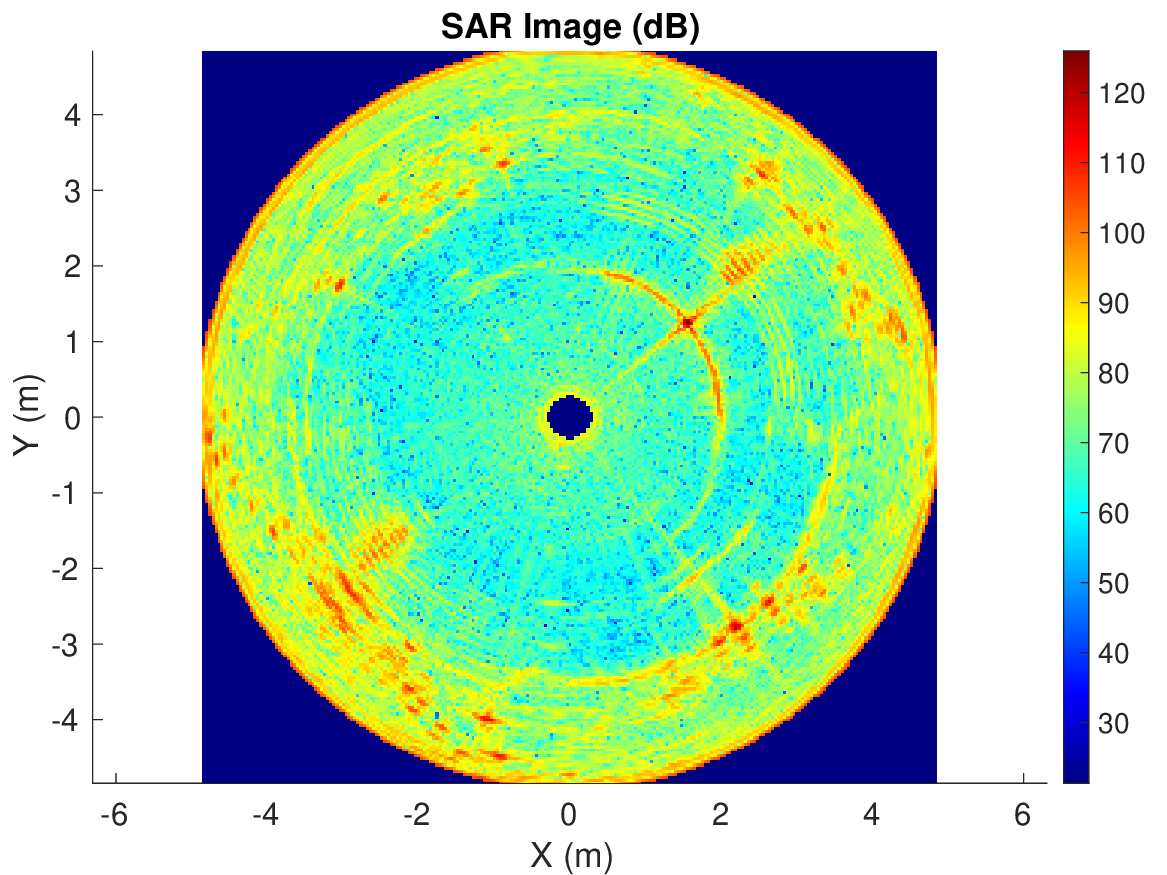}
		\caption{\blue{SAS, \(\phi_{\text{MW}} = 1\degree\), \(\eta = - 30\)dB, Robust}}
	\end{subfigure}
	\begin{subfigure}{0.49\linewidth}
		\includegraphics[width=\linewidth]{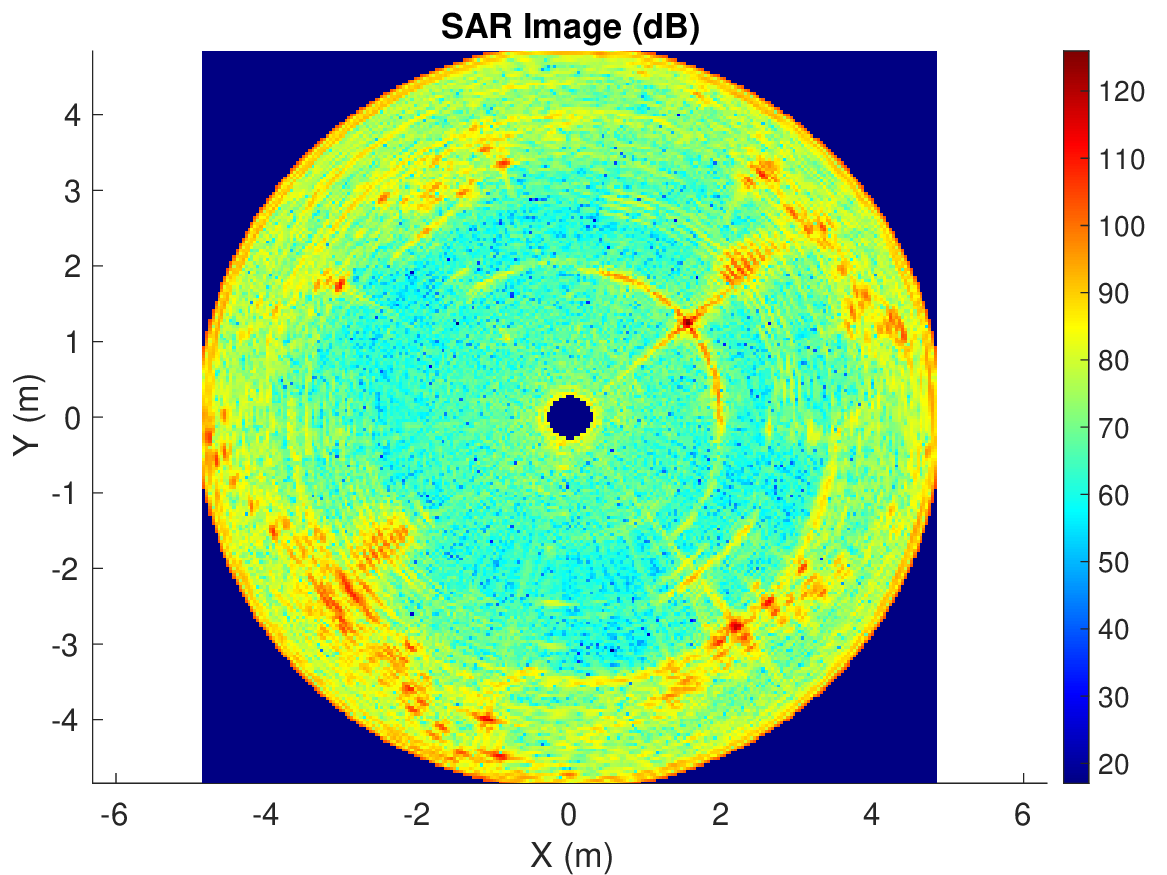}
		\caption{\blue{SAS, \(\phi_{\text{MW}} = 1\degree\), \(\eta = - 33\)dB, Robust}}
	\end{subfigure}
	\caption{SAR Images of a corner reflector (BPA and SAS)}
	\label{fig:SARImages(Time)_CornerReflector}
\end{figure}

\begin{figure}
	\begin{subfigure}{0.49\linewidth}
		\includegraphics[width=\linewidth]{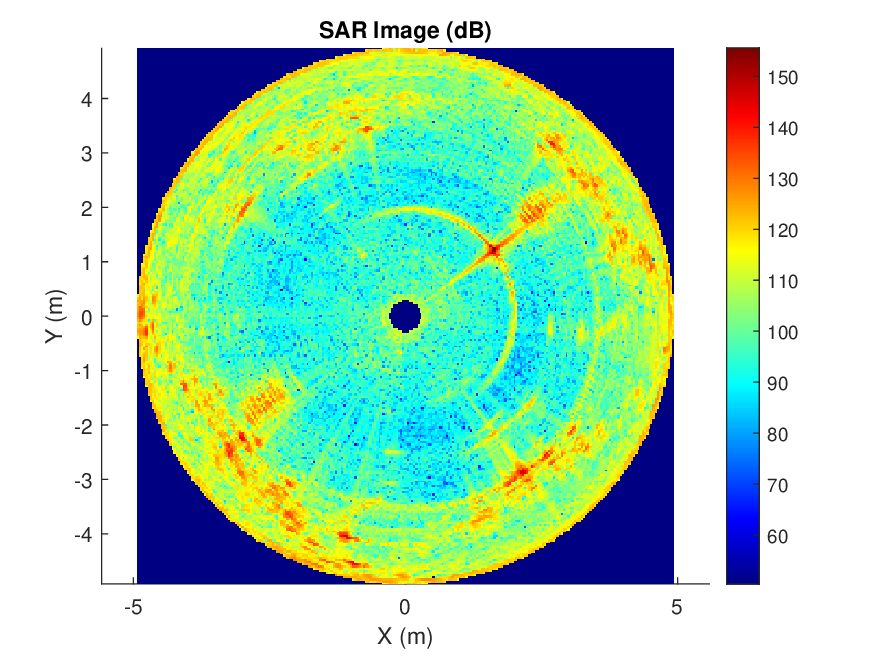}
		\caption{FFT + Back-projection\\algorithm}
	\end{subfigure}
	\begin{subfigure}{0.49\linewidth}
		\includegraphics[width=\linewidth]{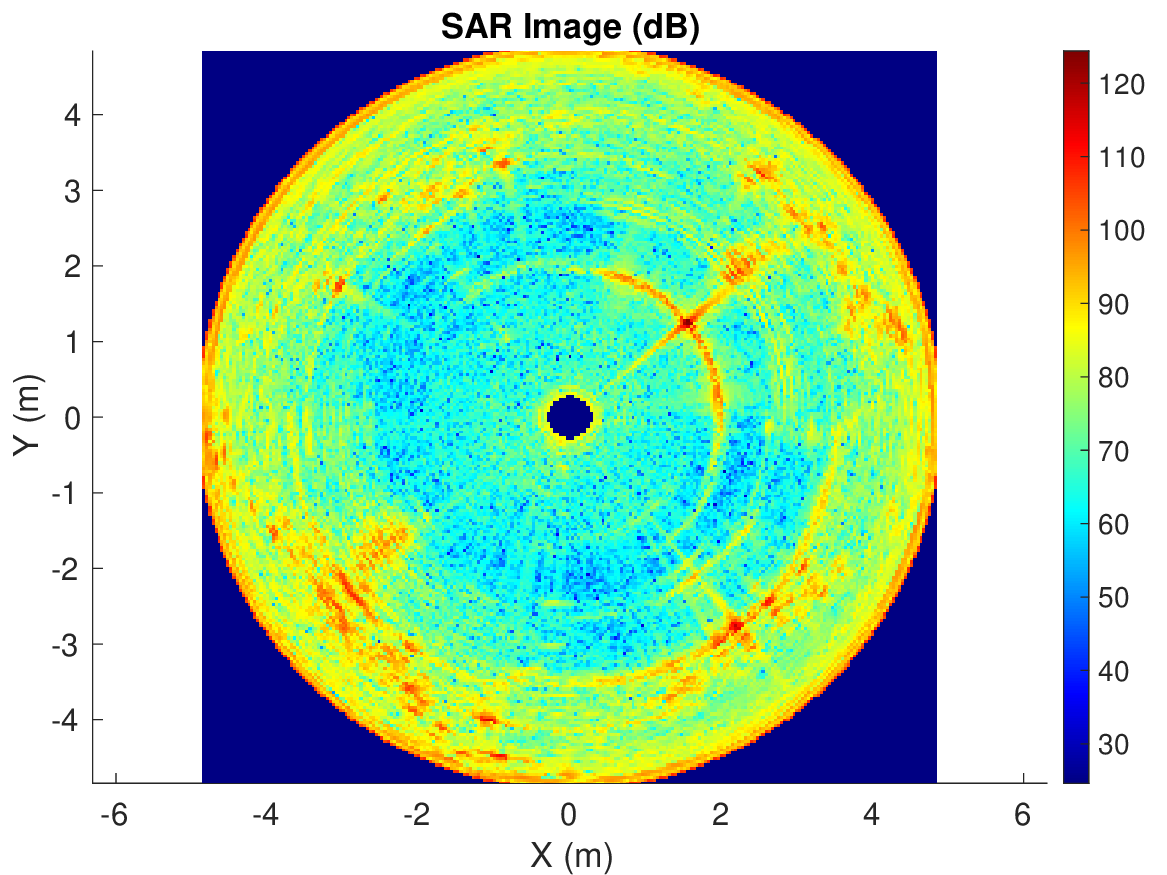}
		\caption{\blue{FFT + SAS, \(\phi_{\text{MW}}\!\!=\!\!1\degree\),\(\eta\!\!=\!\!-30\)dB,w/o Robust}}
	\end{subfigure}
	\begin{subfigure}{0.49\linewidth}
		\includegraphics[width=\linewidth]{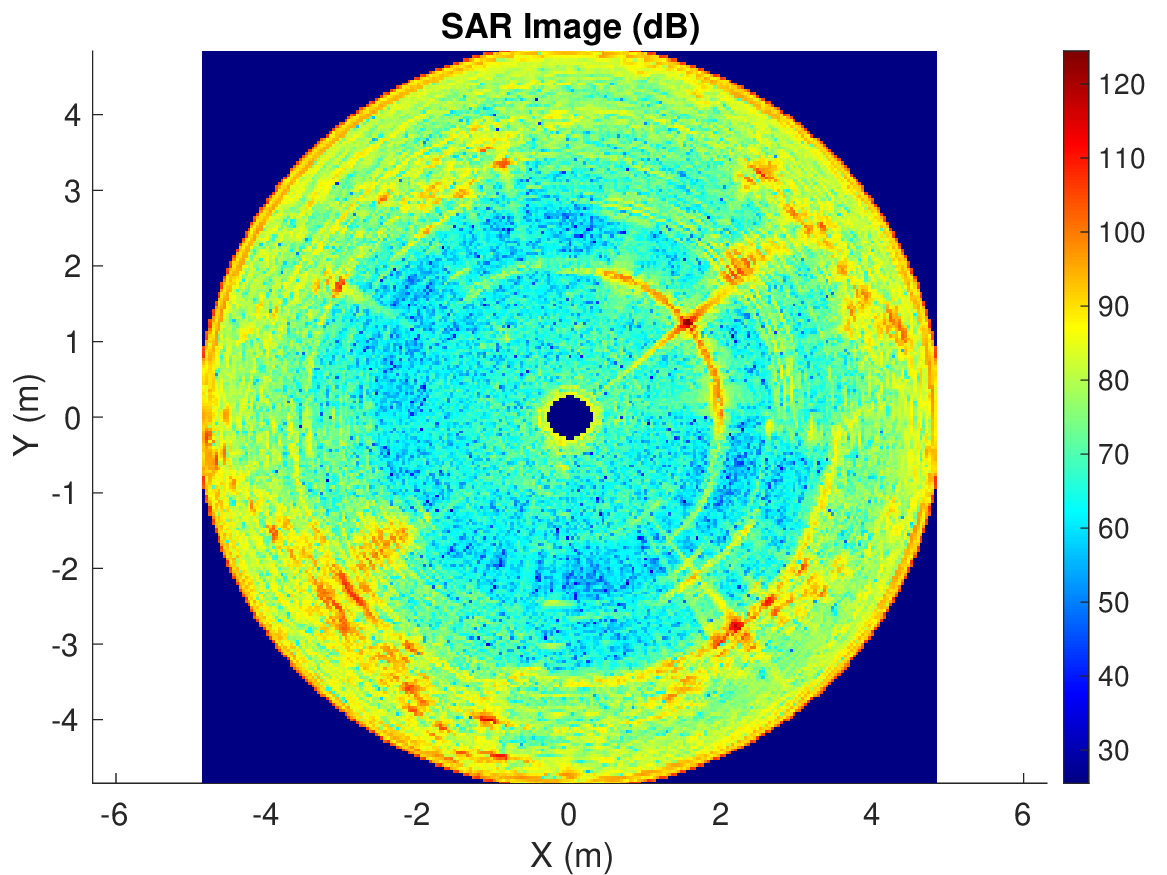}
		\caption{\blue{FFT + SAS, \(\phi_{\text{MW}} = 1\degree\), \(\eta = - 30\)dB, Robust}}
	\end{subfigure}
	\begin{subfigure}{0.49\linewidth}
		\includegraphics[width=\linewidth]{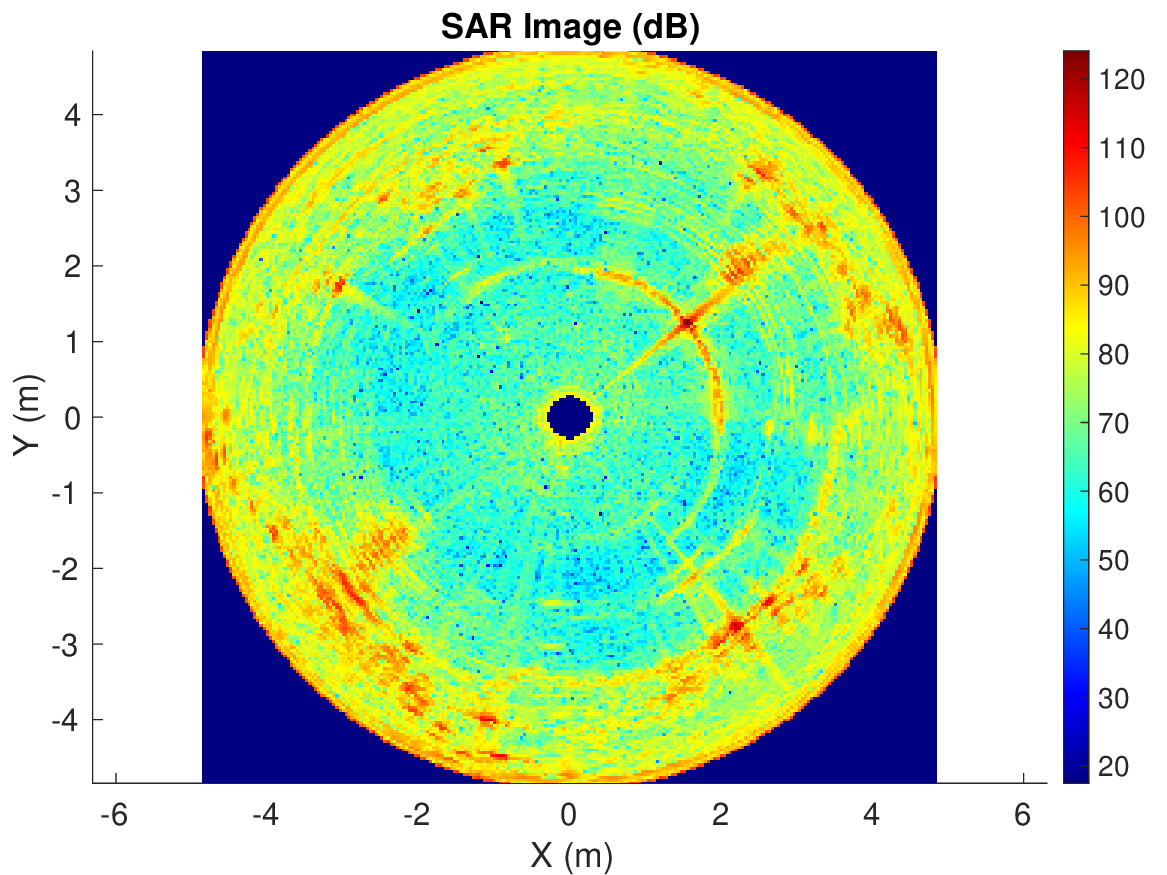}
		\caption{\blue{FFT + SAS, \(\phi_{\text{MW}} = 1\degree\), \(\eta = - 33\)dB, Robust}}
	\end{subfigure}
	\caption{SAR images of a corner reflector (FFT + BPA and FFT + SAS)}
	\label{fig:SARImages(1D)_CornerReflector}
\end{figure}

\begin{figure}
	\begin{subfigure}{0.49\linewidth}
		\includegraphics[width=\linewidth]{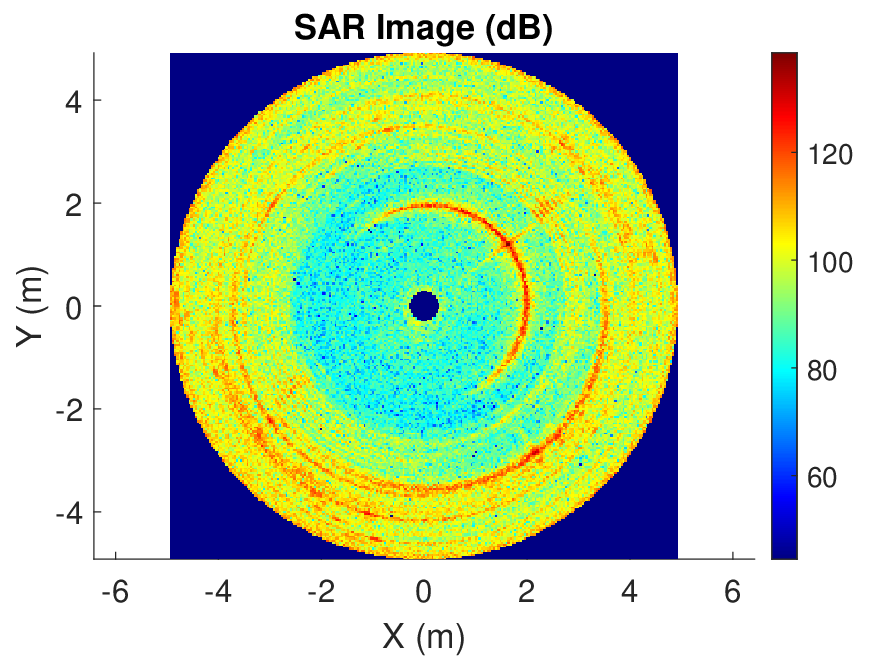}
		\caption{\blue{RBPA}}
	\end{subfigure}
	\begin{subfigure}{0.49\linewidth}
		\includegraphics[width=\linewidth]{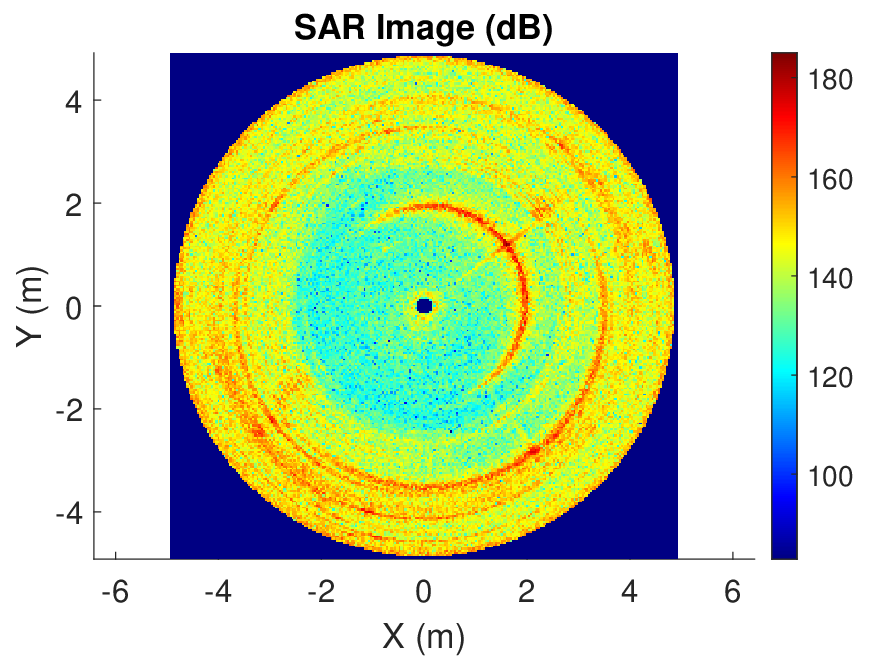}
		\caption{\blue{FFT + RBPA}}
	\end{subfigure}
	\caption{\blue{SAR images of a corner reflector (RBPA and FFT + RBPA)}}
	\label{fig:SARImages(Random)_CornerReflector}
\end{figure}

We put a radar platform (red dot) and a corner reflector (blue dot) in a lab (see Fig. \ref{fig:CornerReflector}). In addition to a corner
reflector which can be treated as the point target with strong reflection in
practice, there are also computer desks, wood cabinet, metal cases and other equipment in the environment. Fig. \ref{fig:SARImages(Time)_CornerReflector}\blue{, \ref{fig:SARImages(1D)_CornerReflector} and \ref{fig:SARImages(Random)_CornerReflector}} illustrate the SAR images under different algorithms and settings. The numerical results are summarized in Table \ref{tab:Algorithmswit DifferentSettings_CornerReflector}. It can be observed from the figures, BPA gives the
clearest image among all approaches. The inclusion of robust design can improve sharpness of the image. At
\(\phi_{\text{MW}} = 1\degree\) and \(\eta = - 33\)dB, the image entropy from SAS is comparable to
that of the BPA but takes only one fifth of the total computation time. Range-FFT can work in conjunction with both BPA and SAS. When comparing \blue{all these figures}, we find that range-dimension match filtering degrades ima{\tiny }ge sharpness slightly. This is also corroborated by the image entropy results in Table \ref{tab:Algorithmswit DifferentSettings_CornerReflector}. Among all approaches,  ``FFT+SAS'' incurs the least amount of compute time -- close to 13 times faster than BPA while achieving acceptable image quality.

\begin{table}
	\centering
	\caption {Imaging Performance with Different Approaches in Corner Reflector Case} \label{tab:Algorithmswit DifferentSettings_CornerReflector} 
	\begin{tabular}{|>{\centering\arraybackslash}m{1.15cm}|>{\centering\arraybackslash}m{3.8cm}|c|>{\centering\arraybackslash}m{1.6cm}|}
		\hline
		Algorithm & Settings & $E_{I}$ & Time Cost (s)\\
		\hline
		BPA & N/A & \textbf{6.0472} & 151.84 \\
		\hline
		FFT+BPA & N/A & 6.3400 & 21.69 \\
		\hline
		\multirow{3}{*}{SAS} & \(\phi_{\text{MW}}\!\!=\!\!1\degree\),\(\eta\!\!=\!\!-30\)dB,w/o Robust & \blue{7.2607} & \blue{28.04} \\
		\cline{2-4}
		& \(\phi_{\text{MW}} = 1\degree\),\(\eta = - 30\)dB,Robust & \blue{7.1717} & \blue{28.94} \\
		\cline{2-4}
		& \(\phi_{\text{MW}} = 1\degree\),\(\eta = - 33\)dB,Robust & \blue{6.8913} & \blue{30.56} \\
		\hline
		\multirow{3}{*}{FFT+SAS} & \(\phi_{\text{MW}}\!\!=\!\!1\degree\),\(\eta\!\!=\!\!-30\)dB,w/o Robust & \blue{7.3989} & \blue{\textbf{11.45}} \\
		\cline{2-4}
		& \(\phi_{\text{MW}} = 1\degree\),\(\eta = - 30\)dB,Robust & \blue{7.3227} & \blue{\textbf{12.14}} \\
		\cline{2-4}
		& \(\phi_{\text{MW}} = 1\degree\),\(\eta = - 33\)dB,Robust & \blue{7.1170} & \blue{\textbf{12.31}} \\
		\hline
		\blue{RBPA} & \blue{N/A} & \blue{8.3060} & \blue{25.02} \\
		\hline
		\blue{FFT+RBPA} & \blue{N/A} & \blue{8.3179} & \blue{13.07} \\
		\hline
	\end{tabular}
\end{table}

\subsubsection{Scenario 2: Corridor Corner} \label{CorridorCorner}
\begin{figure}
	\centering
	\includegraphics[width=0.5\linewidth]{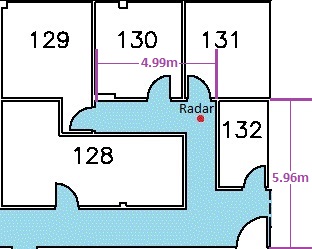}
	\caption{Floor map of the corridor corner}
	\label{fig:CorridorCorner}
\end{figure}

\begin{figure}
	\begin{subfigure}{0.49\linewidth}
		\includegraphics[width=\linewidth]{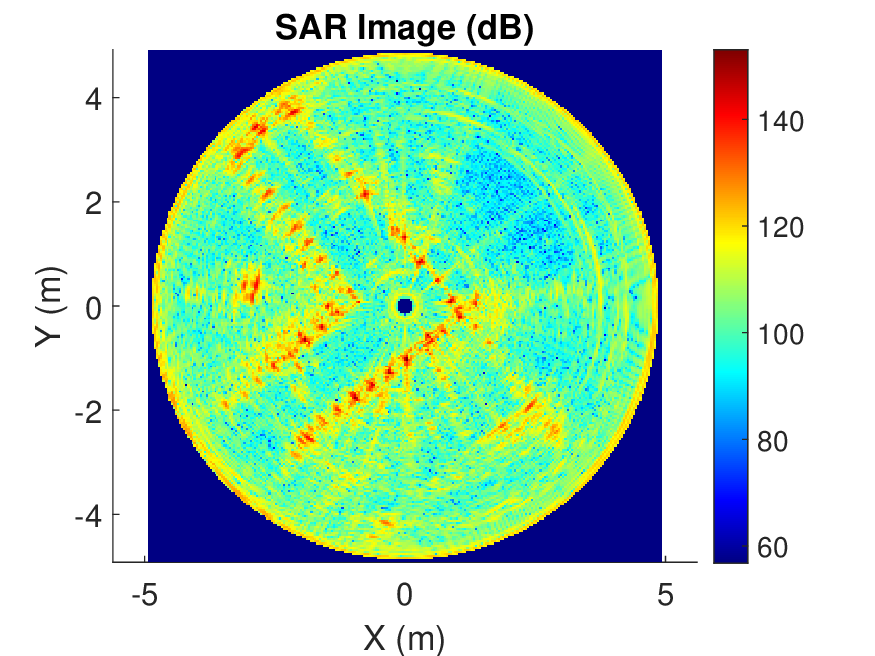}
		\caption{Back-projection\\algorithm}
	\end{subfigure}
	\begin{subfigure}{0.49\linewidth}
		\includegraphics[width=\linewidth]{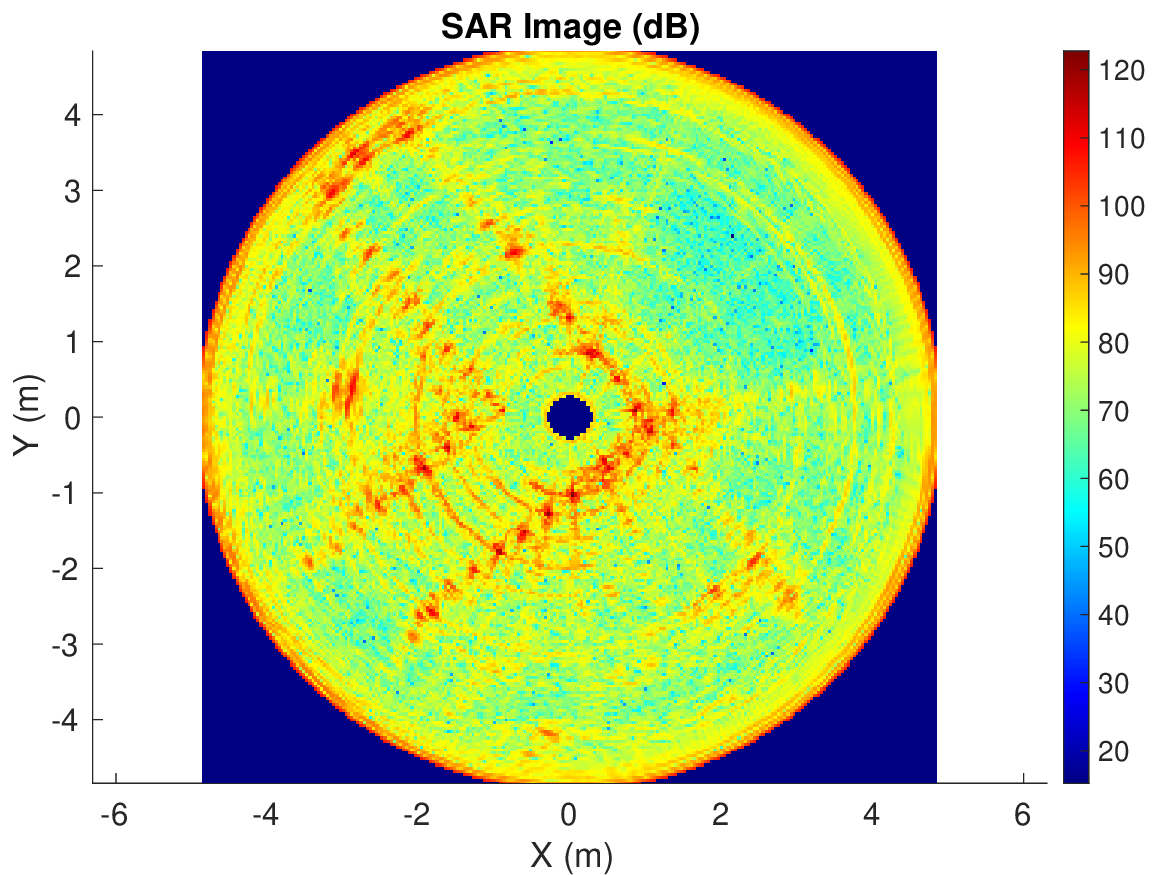}
		\caption{\blue{SAS, \(\phi_{\text{MW}}\!\!=\!\!1\degree\),\(\eta\!\!=\!\!-30\)dB,w/o Robust}}
	\end{subfigure}
	\begin{subfigure}{0.49\linewidth}
		\includegraphics[width=\linewidth]{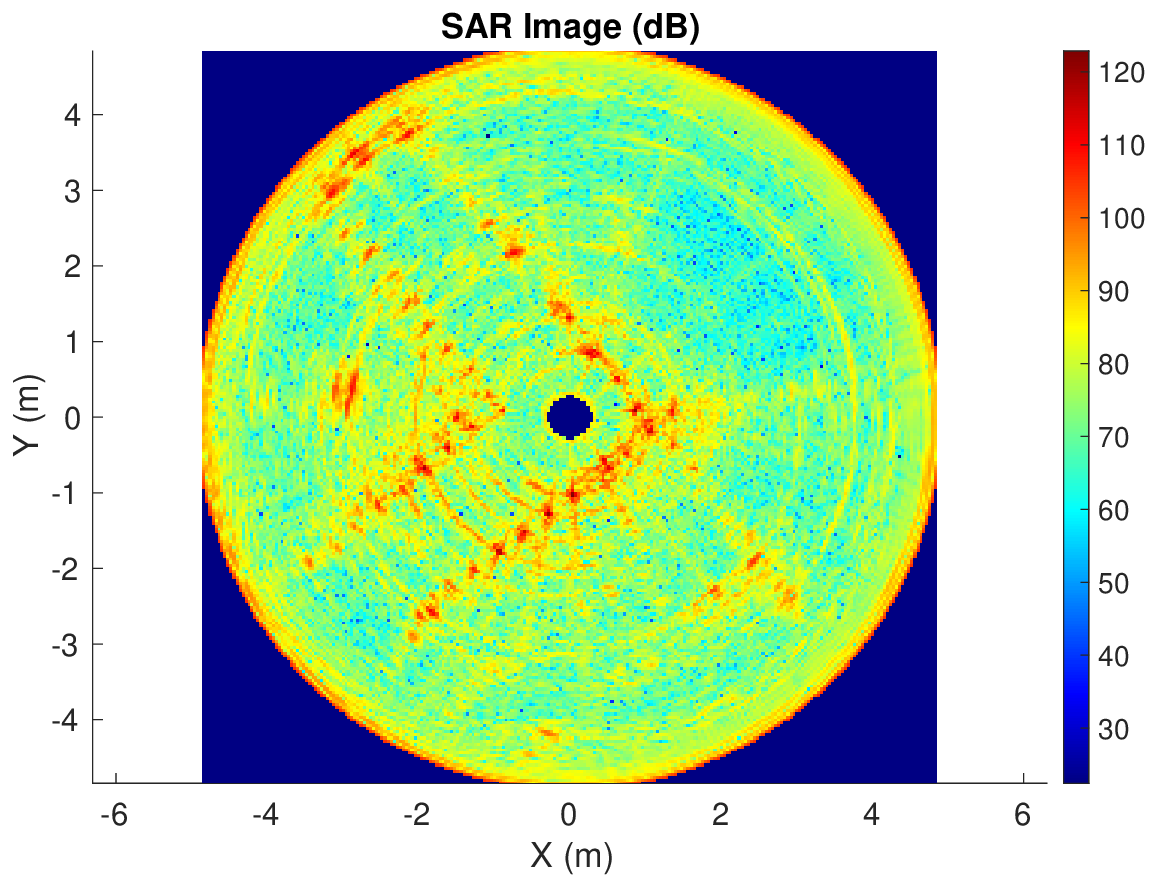}
		\caption{\blue{SAS, \(\phi_{\text{MW}} = 1\degree\), \(\eta = - 30\)dB, Robust}}
	\end{subfigure}
	\begin{subfigure}{0.49\linewidth}
		\includegraphics[width=\linewidth]{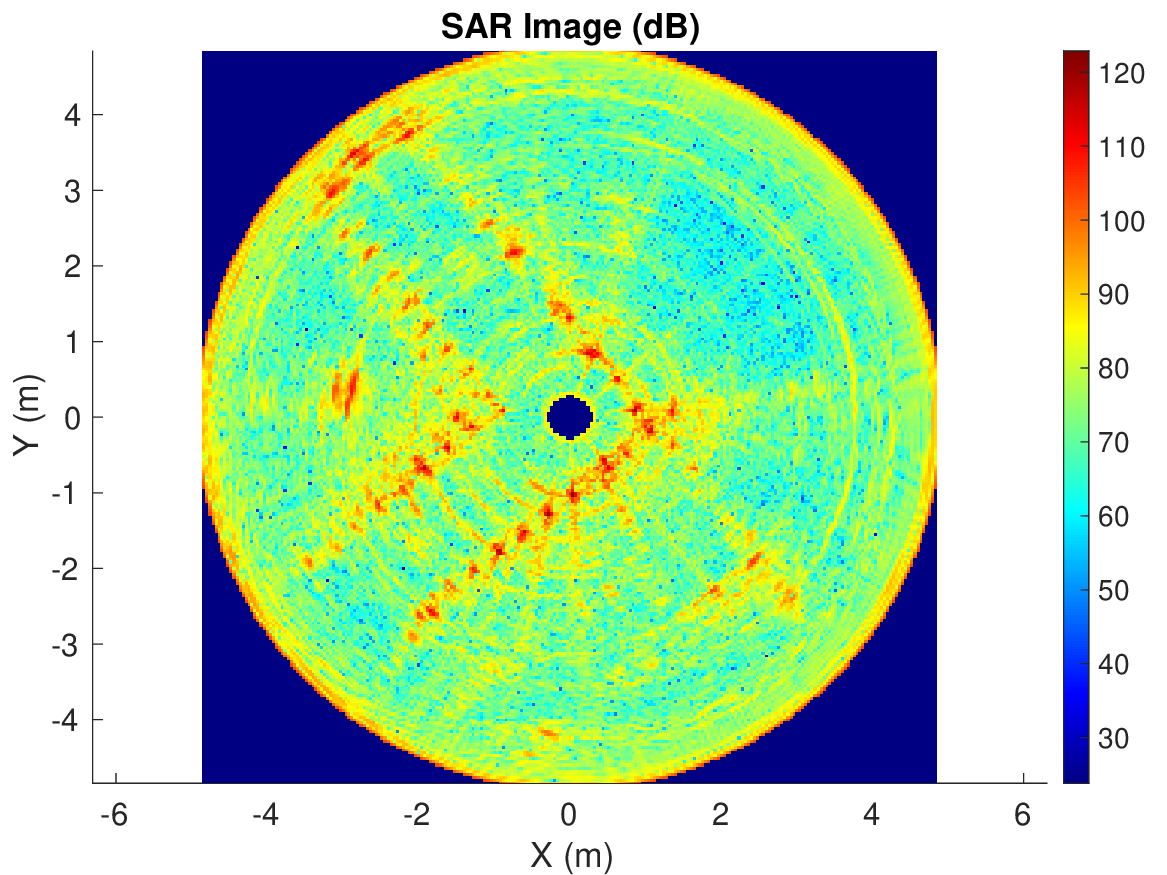}
		\caption{\blue{SAS, \(\phi_{\text{MW}} = 1\degree\), \(\eta = - 33\)dB, Robust}}
	\end{subfigure}
	\caption{SAR images of corridor corner (BPA and SAS)}
	\label{fig:SARImages(Time)_CorridorCorner}
\end{figure}

\begin{figure}
	\begin{subfigure}{0.49\linewidth}
		\includegraphics[width=\linewidth]{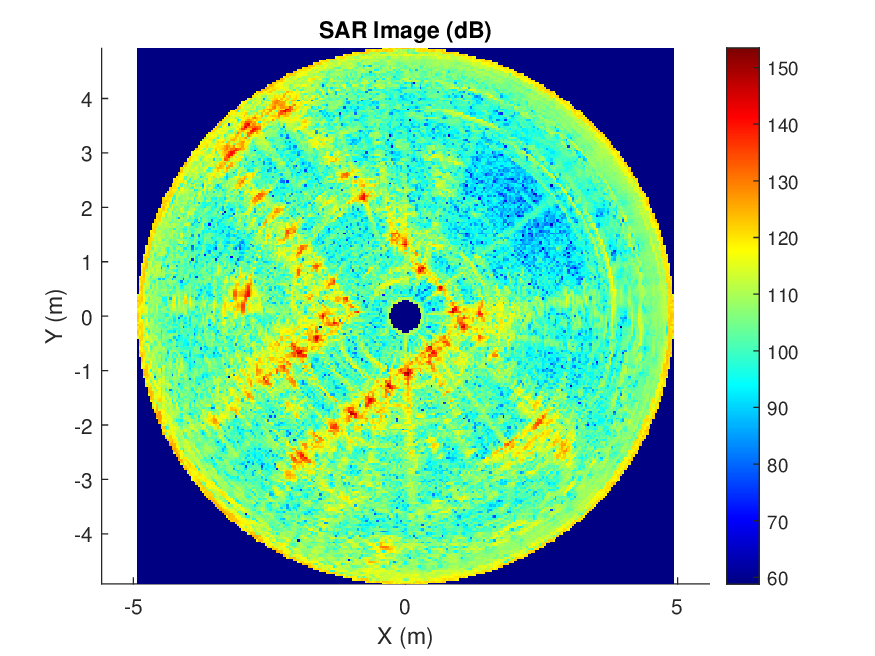}
		\caption{FFT + Back-projection\\algorithm}
	\end{subfigure}
	\begin{subfigure}{0.49\linewidth}
		\includegraphics[width=\linewidth]{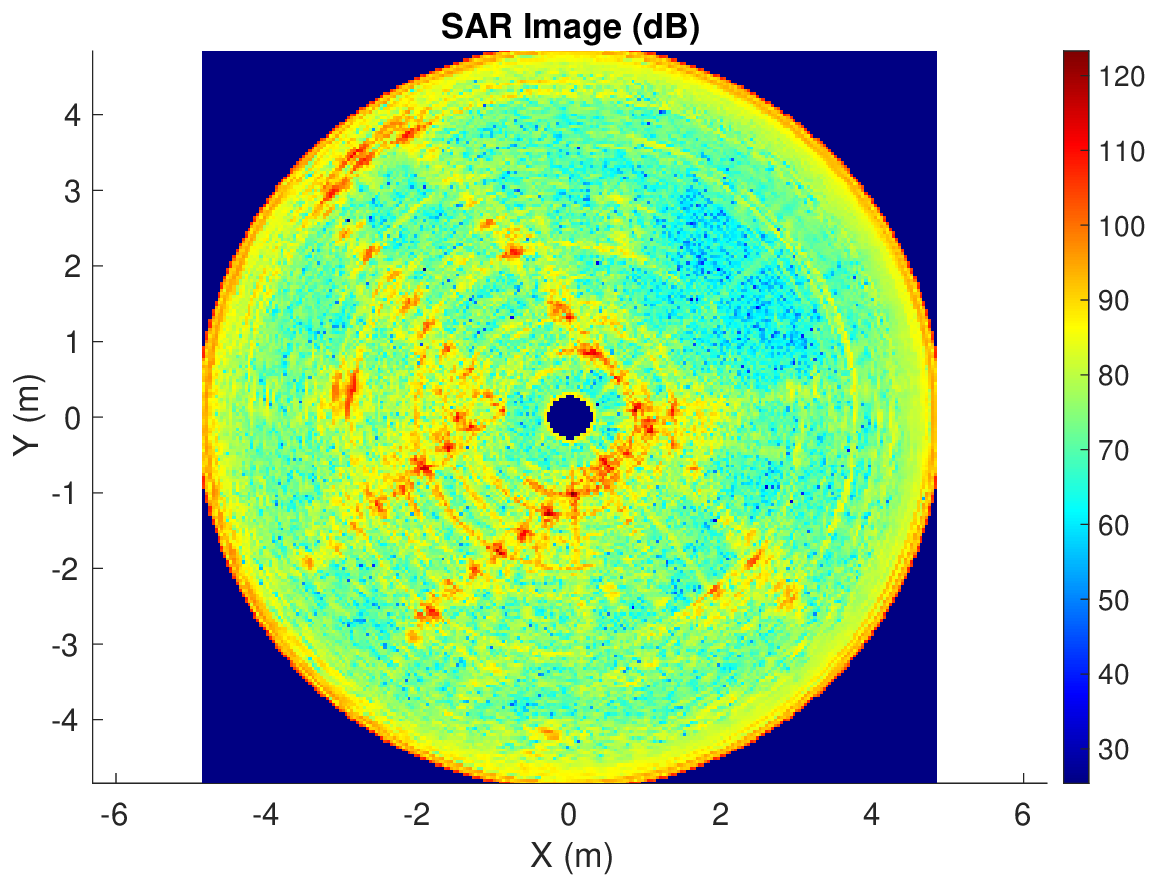}
		\caption{\blue{FFT + SAS, \(\phi_{\text{MW}}\!\!=\!\!1\degree\),\(\eta\!\!=\!\!-30\)dB,w/o Robust}}
	\end{subfigure}
	\begin{subfigure}{0.49\linewidth}
		\includegraphics[width=\linewidth]{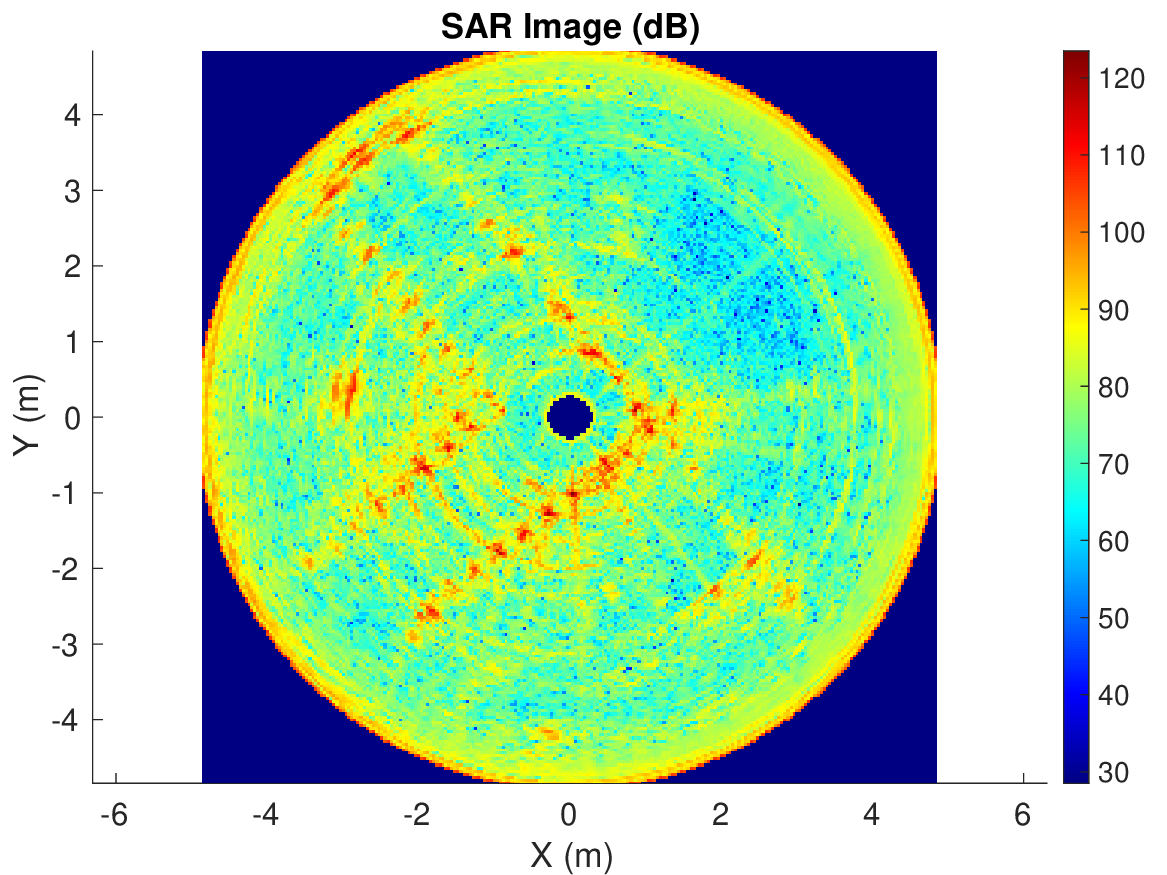}
		\caption{\blue{FFT + SAS, \(\phi_{\text{MW}} = 1\degree\), \(\eta = - 30\)dB, Robust}}
	\end{subfigure}
	\begin{subfigure}{0.49\linewidth}
		\includegraphics[width=\linewidth]{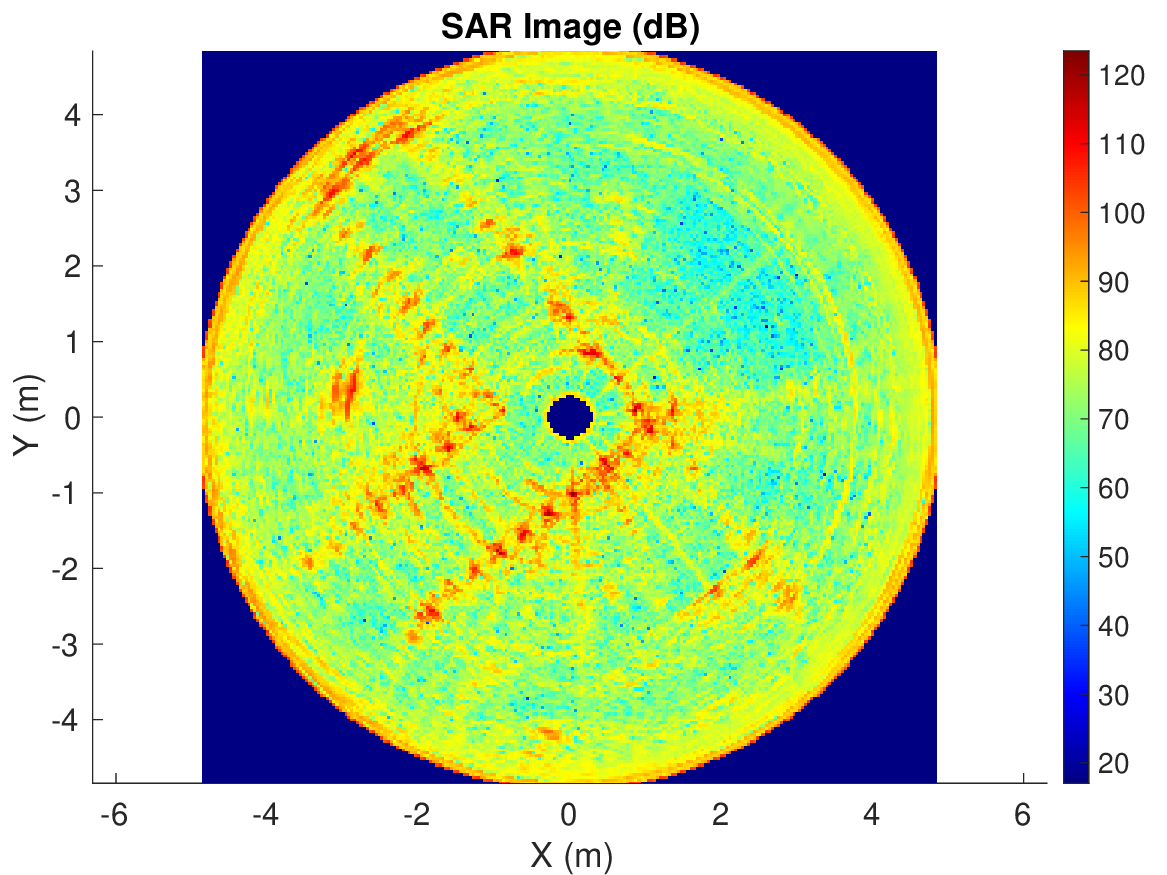}
		\caption{\blue{FFT + SAS, \(\phi_{\text{MW}} = 1\degree\), \(\eta = - 33\)dB, Robust}}
	\end{subfigure}
	\caption{SAR images of corridor corner (FFT + BPA and FFT + SAS)}
	\label{fig:SARImages(1D)_CorridorCorner}
\end{figure}

\begin{figure}
	\begin{subfigure}{0.49\linewidth}
		\includegraphics[width=\linewidth]{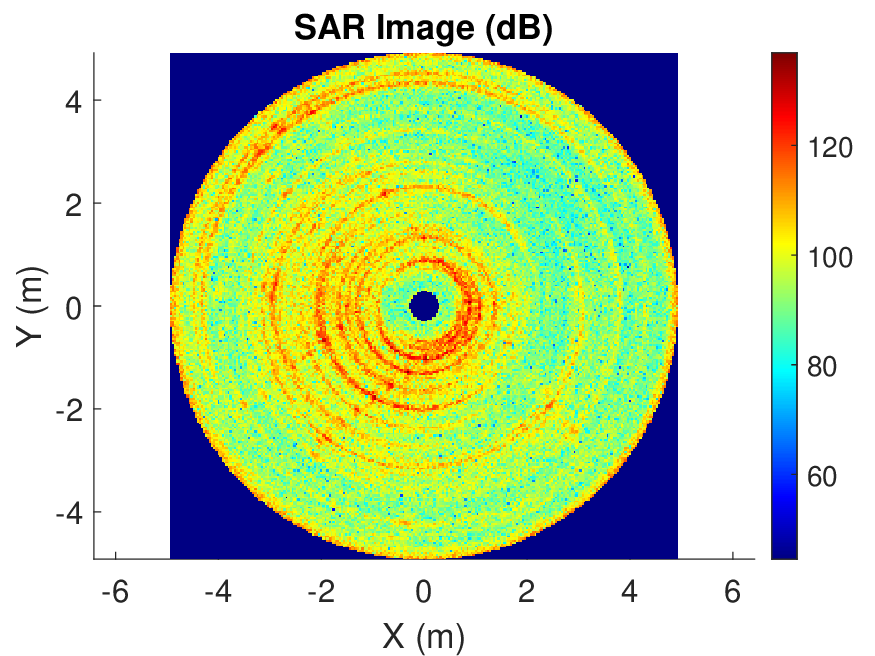}
		\caption{\blue{RBPA}}
	\end{subfigure}
	\begin{subfigure}{0.49\linewidth}
		\includegraphics[width=\linewidth]{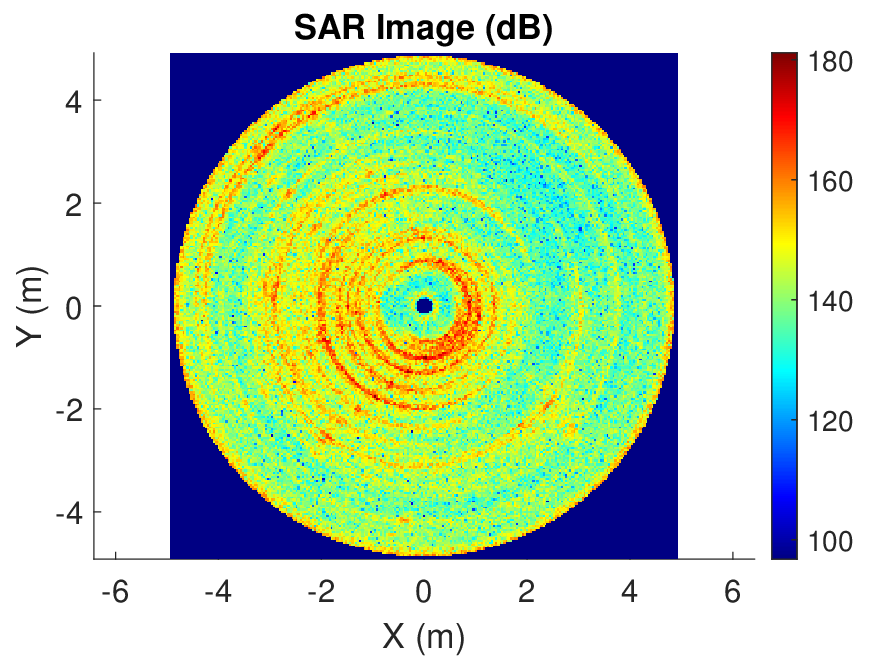}
		\caption{\blue{FFT + RBPA}}
	\end{subfigure}
	\caption{\blue{SAR images of corridor corner (RBPA and FFT + RBPA)}}
	\label{fig:SARImages(Random)_CorridorCorner}
\end{figure}

Next we collect data from the corner of a corridor. The floor map is given in Fig. \ref{fig:CorridorCorner}, where the
red dot indicates the location of the system, the blue area corresponds to the corridor and the 
white space with labels represents different rooms. Figs. \ref{fig:SARImages(Time)_CorridorCorner}, \ref{fig:SARImages(1D)_CorridorCorner} and \ref{fig:SARImages(Random)_CorridorCorner} show the SAR
images from different algorithms and parameter settings. The numerical results are summarized in Table
\ref{tab:Algorithmswit DifferentSettings_CorridorCorner}. From the figures, one can discern the outline of the corridor corner. Due to the penetration of mmWave signals through drywalls, the steel bars inside the walls are visible in the figures. Additionally, an object in the Room 128 and the contour of Room 132 are also visible. Similar to the case with corner reflector, the consideration of robust design can indeed improve image quality. When $\eta$ is set to -33dB and robust design, the sidelobes are less visible. \blue{Moreover, }inclusion of range-domain FFT can indeed greatly reduce the compute time. \blue{BPA with randomly selected phase centers takes less time, but the generated image is blurred.} The proposed robust design, with \(\phi_{\text{MW}} = 1\degree\) and \(\eta = - 33\)dB, gives comparable image quality as that of BPA and consume much less computation time.

\begin{table}
	\centering
	\caption {Imaging Performance with Different Approaches in Corridor Corner Case} \label{tab:Algorithmswit DifferentSettings_CorridorCorner} 
	\begin{tabular}{|>{\centering\arraybackslash}m{1.15cm}|>{\centering\arraybackslash}m{3.8cm}|c|>{\centering\arraybackslash}m{1.6cm}|}
		\hline
		Algorithm & Settings & $E_{I}$ & Time Cost (s)\\
		\hline
		BPA & N/A & \textbf{6.4436} & 151.22 \\
		\hline
		FFT+BPA & N/A & 6.6881 & 21.85 \\
		\hline
		\multirow{3}{*}{SAS} & \(\phi_{\text{MW}}\!\!=\!\!1\degree\),\(\eta\!\!=\!\!-30\)dB,w/o Robust & \blue{7.3802} & \blue{27.75} \\
		\cline{2-4}
		& $\phi_{\text{MW}} = 1\degree$,$\eta = - 30$dB,Robust & \blue{7.2871} & \blue{29.30} \\
		\cline{2-4}
		& $\phi_{\text{MW}} = 1\degree$,$\eta = - 33$dB,Robust & \blue{7.0251} & \blue{30.25} \\
		\hline
		\multirow{3}{*}{FFT+SAS} & \(\phi_{\text{MW}}\!\!=\!\!1\degree\),\(\eta\!\!=\!\!-30\)dB,w/o Robust & \blue{7.5309} & \blue{\textbf{12.50}} \\
		\cline{2-4}
		& $\phi_{\text{MW}} = 1\degree$,$\eta = - 30$dB,Robust & \blue{7.4562} & \blue{\textbf{12.21}} \\
		\cline{2-4}
		& $\phi_{\text{MW}} = 1\degree$,$\eta = - 33$dB,Robust & \blue{7.2425} & \blue{\textbf{12.47}} \\
		\hline
		\blue{RBPA} & \blue{N/A} & \blue{8.3671} & \blue{25.35} \\
		\hline
		\blue{FFT+RBPA} & \blue{N/A} & \blue{8.3801} & \blue{13.28} \\
		\hline
	\end{tabular}
\end{table}

\section{Conclusion} \label{sec:Conclusion}
In this paper, we propose a new fast imaging algorithm based on robust sparse array synthesis for ROSAR. Since radar path is circular, such an algorithm only needs to pre-compute the complex weights of the imaging filter offline for one direction per range bin. Due to the external influence could affect the sidelobe level, we add robust design to maintain the image quality. To meet our pre-set expectation and solve this problem, our proposed algorithm employs feasible point pursuit and successive convex approximation technology. On that basis, we also give another algorithm based on range-FFT to further reduce the computation complexity.

According to the simulation and testbed results, we conclude that our approach can generate an SAR image with the quality comparable to that of BPA. Meanwhile, the proposed approach is able to reduce the computational cost significantly and is robust to the array error.

Nonetheless, we must sacrifice some image quality if employing FFT-based range-dimensional matched filtering. Thus, exploring a better approach for the processing based on range FFT is our future research direction.

\bibliographystyle{IEEEtran}
\bibliography{IEEEabrv,Reference}

\vskip -3.5\baselineskip plus -1fil
\textbf{\begin{IEEEbiographynophoto}{Wei Zhao} received his B.Eng. degree in Automation from Xidian University, M.A.Sc. degree in Electrical \& Computer Engineering from McMaster University. He is a Ph.D. candidate in the Department of Computing and Software in McMaster University since 2017.\end{IEEEbiographynophoto}}
\vskip -3.5\baselineskip plus -1fil
\textbf{\begin{IEEEbiographynophoto}{Cai Wen} (M’19) received the B.E. degree from the School of Electronic Engineering, Xidian University, Xi’an, China, in 2009, and the Ph.D. degree from the National Laboratory of Radar Signal Processing, Xidian University, Xi’an, China, in 2014. He was a Research Scientist with the China NORINCO Group from Jan. 2015 to Oct. 2016. From Nov. 2019 to Mar. 2023, he was with the Department of Electrical and Computer Engineering, McMaster University, Hamilton, Canada, as a Post-Doctoral Research Fellow. Since Nov. 2016 he has been with the School of Information Science and Technology, Northwest University, Xi’an, China, where he is currently an Associate Professor. His current research interests include sensor array signal processing, MIMO radar signal processing, integrated sensing and communication, and mathematical optimization.\end{IEEEbiographynophoto}}
\vskip -3.5\baselineskip plus -1fil
\textbf{\begin{IEEEbiographynophoto}{Quan Yuan}	received the Bachelor of Engineering degree from McMaster University, Canada, in 2022. He is currently working towards the Master of Engineering degree on Computing and Software under the supervision of Dr. Rong Zheng at McMaster University, Canada.\end{IEEEbiographynophoto}}
\vskip -3.5\baselineskip plus -1fil
\textbf{\begin{IEEEbiographynophoto}{Rong Zheng} received the B.E. and M.E. degrees in electrical engineering from Tsinghua University, Beijing, China, in 1996 and 1998, respectively, and the Ph.D. degree from the Department of Computer Science, University of Illinois at Urbana–Champaign, Champaign, IL, USA, in 2004. She is currently a Professor with the Department of Computing and Software, McMaster University, Hamilton, ON, Canada.\end{IEEEbiographynophoto}}
\end{document}